\pdfoutput=1
\documentclass[numbers,final,3p,times]{elsarticle}
\usepackage{amssymb}
\usepackage{amsthm}
\usepackage{amsmath}
\usepackage{bm}
\usepackage{array}
\usepackage{ulem}
\usepackage{longtable}
\usepackage{setspace}
\usepackage{multirow}
\usepackage{siunitx}
\usepackage{threeparttable}
\usepackage{verbatim}
\usepackage{color}
\usepackage[justification=centering]{caption}
\usepackage{lineno}

\begin{document}
\begin{spacing}{1.0}
\begin{frontmatter}


\title{Direct reduction of iron-ore with hydrogen in fluidized beds: A coarse-grained CFD-DEM-IBM study}

\author[label1,label2]{Bin Lan}
\author[label2,label3]{Ji Xu}
\author[label2,label3]{Shuai Lu}
\author[label2,label3]{Yige Liu}
\author[label1,label2]{Fan Xu}
\author[label2,label3]{Bidan Zhao}
\author[label2,label3]{Zheng Zou}
\author[label1]{Ming Zhai\corref{cor1}}
\ead{zhaiming@hit.edu.cn}
\author[label2,label3,label4]{Junwu Wang\corref{cor1}}
\cortext[cor1]{Corresponding author}
\ead{jwwang@ipe.ac.cn}

\address[label1]{School of Energy Science and Engineering, Harbin Institute of Technology, Harbin, 150001, P. R. China}
\address[label2]{State Key Laboratory of Multiphase Complex Systems, Institute of Process Engineering, Chinese Academy of Sciences, P. O. Box 353, Beijing 100190, P. R. China}
\address[label3]{School of Chemical Engineering, University of Chinese Academy of Sciences, Beijing, 100049, P. R. China}
\address[label4]{Innovation Academy for Green Manufacture, Chinese Academy of Sciences, Beijing, 100190, P. R. China}

\begin{abstract}
Hydrogen metallurgy technology uses hydrogen as the reducing agent instead of carbon reduction, which is one of the important ways to reduce carbon dioxide emissions and ensure the green and sustainable development of iron and steel industry. Due to the advantages of high gas-solid contact efficiency and outstanding mass and heat transfer, direct reduction of iron ore in fluidized beds has attracted much attention. In this study, a coarse-grained CFD-DEM-IBM solver based on hybrid CPU-GPU computing is developed to simulate the direct reduction process of two kinds of iron ore with hydrogen in fluidized beds, where an unreacted shrinking core model based on multiple reaction paths is used to model the reduction reactions, a coarse-grained model and multiple GPUs enable the significant acceleration of particle computation, and the immersed boundary method (IBM) enables the use of simple mesh even in complex geometries of reactors. The predicted results of particle reduction degree are in good agreement with the experimental values, which proves the correctness of the CFD-DEM-IBM solver. In addition, the effects of reaction kinetic parameters and operating temperature on particle reduction degree are also investigated. Present study provides a method for digital design, optimization and scale-up of ironmaking reactors.
\end{abstract}

\begin{keyword}
Direct reduction; Fluidization; CFD-DEM; Unreacted shrinking core model; Coarse-graining
\end{keyword}
\end{frontmatter}

\section{Introduction}\label{s1}
The steel industry has played a great role in supporting and promoting national defense, petroleum, shipbuilding, buildings, and equipment manufacturing \citep{zhou2016emission}. Metallic iron, as one of the main components of steel, can generally be produced by three process routes \citep{dutta2020basic}: blast furnace, direct reduction and smelting reduction. At present, due to the advantages of mature technology and high output, blast furnace ironmaking is the dominant process in steel production \citep{soni2022review}. However, the production process of blast furnace ironmaking is complicated, and the carbon dioxide and sulfur dioxide emissions generated after the combustion of the coke used in smelting are very high, which causes serious pollution to the environment \citep{lu2015quality}. In order to alleviate the pressure of ironmaking on environment, resource and energy, non-blast furnace technology has been paid more and more attention by researchers \citep{sun2014gas,yi2019finex,jeong2015system}. Non-blast furnace ironmaking completely get rid of the dependence on coke and realize the comprehensive utilization of complex symbiotic iron ore resources while meeting the requirements of environmental protection.

As a typical non-blast furnace ironmaking method, fluidized ironmaking directly uses pulverized ore as raw material, which eliminates the process of pelletizing and sintering and helps to realize the efficient utilization of low-grade complex iron ore \citep{kwauk2007handbook}. The direct reduction of iron ore in fluidized beds is a promising method for iron making because of its high heat and mass transfer efficiency and rapid reduction rate \citep{he2017direct}. The world's first Fluid Iron Ore Reduction (FIOR) process appeared in 1960 and was commercially operated in Venezuela in 1976 until it was shut down in 2000 \citep{brown1966fior}. After that, various iron and steel enterprises successively established FINMET, FINEX, Circored, Circofer and other processes \citep{battle2014direct}. These processes have one thing in common, that is, the fluidized reduction system is generally composed of two or more fluidized bed reactors in series, and the gas and particles are in countercurrent contacting state. The advantages of this design lie in the fact that it can make full use of the waste heat of reducing gas to preheat iron ore powder to speed up the reaction rate, to reduce the consumption of reducing gas and to improve the gas utilization rate \citep{plaul2009fluidized}. Many researchers have conducted experimental studies on the hydrodynamics and reaction mechanism in fluidized bed for iron ore reduction \citep{yu2010application,yu2019growth,li2015phase,du2016role,du2022effect,du2022relationship,du2017enhanced,spreitzer2019iron,spreitzer2020fluidization}. It was found that fine iron ore powder (about 100 $\mu$m) is easy to stick together when reduced at high temperature \citep{zhong2018model}, thus forming clusters and depositing at the bottom of the bed, and finally resulting in the occurrence of defluidization. Adhesion and defluidization can be effectively inhibited by granulation \citep{du2022relationship}, solid additive coating \citep{komatina2018sticking}, carbon deposition-reduction \citep{lei2016optimization}, reducing reduction temperature \citep{lei2014experimental}, and/or increasing operating gas velocity \citep{zhong2012defluidization}. The use of pure hydrogen as a reducing agent to make iron is known as hydrogen metallurgy, and this technology has attracted much attention in recent years due to its ability to achieve $\rm{CO_2}$ emission reduction targets \citep{holappa2017recent,an2018potential,patisson2020hydrogen}.

In addition to experiment, numerical simulation has gradually become an important means to study the direct reduction of iron ore \citep{shi2005modelling,valipour2009mathematical,tang2012simulation,natsui2014numerical,nouri2011simulation,ariyan2016numerical,
kinaci2018direct,kinaci2020cfd,schneiderbauer2020computational,wan2022numerical,rosser2023investigation}. Among many numerical calculation methods for reduction reaction, CFD-DEM method is popular. The DEM model can provide information of force, velocity, position, temperature, reaction rate and other important data on the particle scale, which are difficult to obtain by experiment. Therefore, DEM simulation is extremely helpful for researchers and engineers to understand the hydrodynamics, heat and mass transfer as well as chemical reaction characteristics of particles in the reactor. Natsui \cite{natsui2014numerical} adopted CFD-DEM method to simulate the process of direct reduction of large hematite particles (with a particle size of 4 cm) to iron in three-dimensional packed bed, and observed that the heterogeneity of the reaction rate and the temperature distribution was affected by particle arrangement.
Dianyu \cite{dianyu2020validation} found that with the increase of operating pressure (below 5 atm), the reduction rate of single hematite pellet increased significantly, and the entire reduction process under non-isothermal conditions was slower than that under isothermal conditions. Xu \cite{xu2022coarse} used a coarsed-grained CFD-DEM model to study the simplified reduction reaction of iron ore accompanied by coke gasification in blast furnaces. In the gas-solid chemical reaction, the particles may shrinking or expansion, and their surfaces may form an outer layer of solid products. Shrinkage/expansion in a reaction is often described by the shrinking core model (SCM) and the unreacted core model (UCM) \citep{levenspiel1998chemical}. In iron ore reduction reactions, these two phenomena complement each other and are represented by the unreacted shrinking core model (USCM) \citep{valipour2009mathematical}. Kinaci \cite{kinaci2017modelling,kinaci2018direct,kinaci2020cfd} combined the USCM reaction rate model with the CFD-DEM method, and simulated the reduction reaction of centimeter-level iron ore particles.

Although the USCM model has been widely used in numerical studies of reaction flows, a complete resistance network (reaction path) that considers the effects of particle structure, reduction gas concentration, and reaction temperature during the reaction has not been found in the literature. The reduction resistance network is directly related to the reaction resistance and thus affects the reaction rate. Therefore, the accurate construction of mathematical and physical models of iron ore direct reduction is the key to predict the hydrodynamics, mass transfer and heat transfer behavior of gas and solid in the reactor based on CFD-DEM method. In this study, the USCM model is coupled to a coarse-grained CFD-DEM method, where the motions of gas phase and of iron ore particles are described by the continuum model and the soft sphere model, respectively. Immersed boundary method (IBM), which enables the use of Cartesian meshes for complex geometries of reactors, is coupled to CFD-DEM method. This can conveniently implement the boundary conditions of velocity, pressure, temperature and gas composition. At the same time, the pseudo-turbulence models for hydrodynamic, heat and mass transfer are closed using correlations that are generated from particle-resolved direct numerical simulations. In addition to the application of coarse-grained method, the calculation of particles is GPU-based, and the calculation speed of solid phase is significantly accelerated. The kinetic parameter of the reaction (pre-exponential factor) is determined by means of experimental data correction rather than directly from the literature. The reduction process of two different kinds of iron ore in bubbling bed is simulated, and the effects of kinetic parameters and operating temperature on the reduction degree, reaction resistance and porosity of particles are investigated. The rest of the article is organized as follows: Section 2 introduces the reaction model. The CFD-DEM governing equations and constitutive relations are given in Section 3. Section 4 shows numerical method and simulation setup. In Section 5, the simulation results are analyzed and the reduction degree are compared with the experimental data available in literature. The conclusions are drawn in Section 6.

\section{Reaction model}\label{s2}
\begin{figure}
\centerline{\includegraphics[width=0.5\textwidth]{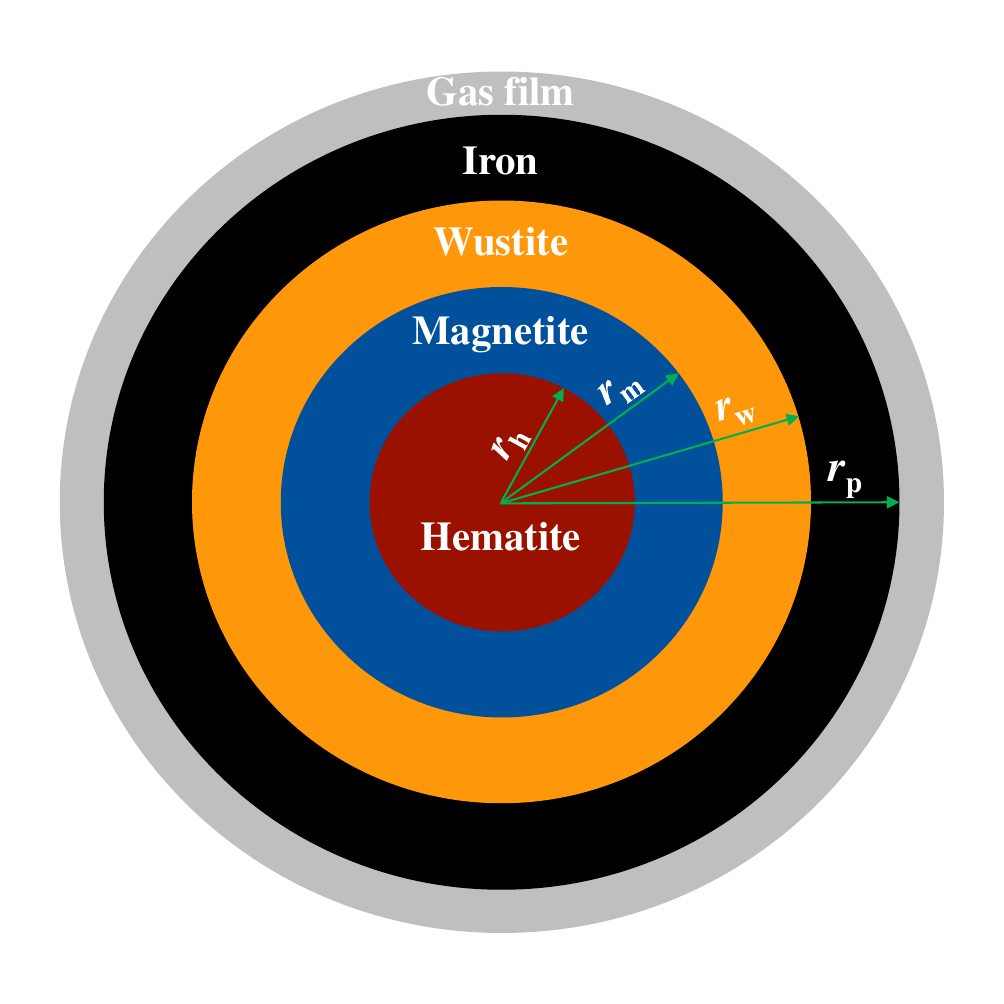}}
\caption{Schematic illustration of three-layer USCM ($r_\text{p}$ is particle radius, $r_\text{h}$, $r_\text{m}$ and $r_\text{w}$ are the radius of core, product layer m and w, respectively).}\label{fig1}
\end{figure}

\begin{figure}
\centerline{\includegraphics[width=0.8\textwidth]{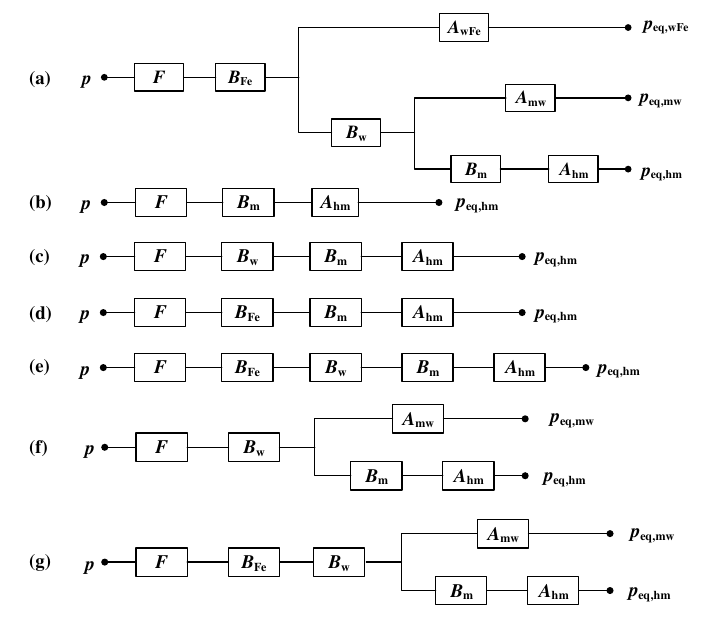}}
\caption{Resistance network for progressive reduction of hematite ($p$ is the gas pressure, $F$ represents the mass transfer resistance through the gas film layer, $B$ represents the diffusion resistance through the product layer and $A$ represents the chemical reaction resistance at the reaction interface. Furthermore, h, m and w represent hematite, magnetite and wustite, respectively).}\label{fig2}
\end{figure}

Hematite (${\rm{Fe}_2\rm{O}_3}$) can be gradually reduced to magnetite (${\rm{Fe}_3\rm{O}_4}$), wustite (FeO) and iron (Fe) in $\rm{H}_2$ atmosphere when the reaction temperature is above 570 $^\circ \text{C}$ \citep{spreitzer2019reduction}. It can be seen from Table \ref{tab1} that the reaction of $\rm{Fe_2O_3}$ to $\rm{Fe_3O_4}$ in the hydrogen atmosphere is exothermic, $\rm{Fe_3O_4}$ to FeO and FeO to Fe are endothermic, so continuous heat input is required to ensure the normal reduction of hematite.
In the three-layer USCM, the particles consist of unreacted hematite core, product layers (magnetite, wustite and metallic iron), and gas film (Figure \ref{fig1}). Based on this model, the following assumptions are made: (i) Each sub-reaction is a first order irreversible reaction; (ii) The radius of the particle does not change, but the structure changes during the whole reduction process; (iii) The sintering effect is not considered, and the structural changes are only caused by chemical reactions; (iv) The temperature inside the particle is uniform because the particles are small; (v) Catalysis and side reactions are ignored. The reduction process of iron ore is considered to be a resistance network \citep{spitzer1966generalized} (Figure \ref{fig2}), which represents three rate control steps, namely, mass transfer resistance $F$ through the gas film layer, diffusion resistance $B$ through the product layer and chemical reaction resistance $A$ at the reaction interface. The mass change rate of the reacting gas can be expressed by the resistance mentioned above by using the Kirchhoff's law \cite{spitzer1966generalized}.

\begin{table}[]
\centering
\caption{Hematite reduction by steps \citep{zhang2013thermodynamic}.}\label{tab1}
\begin{tabular}{cccc}
\hline
Reaction                    & $\Delta H$    & Equation       \\ \hline
${3\rm{Fe}_2\rm{O}_3} + \rm{H_2} \rightarrow{2\rm{Fe}_3\rm{O}_4}+\rm{H_2O}$       &$<$0       & R1            \\
${\rm{Fe}_3\rm{O}_4} + \rm{H_2} \rightarrow{3\rm{Fe}\rm{O}} + \rm{H_2O}$       &$>$0       & R2            \\
${\rm{Fe}\rm{O}} + \rm{H_2} \rightarrow{\rm{Fe}} + \rm{H_2O}$       &$>$0       & R3            \\   \hline
\end{tabular}
\end{table}

\subsection{Reaction rate model: the raw iron ore is hematite}
The rate of mass change of reducing gas ($\text{H}_2$) in reaction $k$$\rightarrow$$l$ is \citep{valipour2009mathematical}:
\begin{equation}\label{e2-1}
\dot m_j^{kl} = \frac{{\eta _j^{kl}{A_{pi}}}}{{{T_{pi}}}}\frac{{{M_j}{p_{\rm{g}}}}}{{{R_{\rm{g}}}}}{\left[ {v_1^{kl}({x_j} - x_{{\rm{eq}},j}^{{\rm{hm}}}) + v_2^{kl}({x_j} - x_{{\rm{eq}},j}^{{\rm{mw}}}) + v_3^{kl}({x_j} - x_{{\rm{eq}},j}^{{\rm{wFe}}})} \right]_j},
\end{equation}
where $k$ and $l$ represent solid reactant ($\rm{Fe_2O_3}$, $\rm{Fe_3O_4}$ or FeO) and product ($\rm{Fe_3O_4}$, FeO or Fe) in each sub-reaction, respectively. Subscript $j$ represents reducing gas, and $A_{pi}$ and $T_{pi}$ are the surface area and temperature of particle $pi$, respectively. $p_\text{g}$ is the gas pressure of the grid where the particle $pi$ is located, and $R_\text{g}$ is the gas constant (8.314 J/mol/K). $M_\text{j}$ and $\eta_\text{j}$ are the molar mass and stoichiometric coefficients of the gas, respectively. Note that for $\rm{H_\text{2}}$, the mole fraction ${x_{{{\rm{H}}_2}}} = {x'_{{{\rm{H}}_2}}}/({x'_{{{\rm{H}}_2}}} + {x'_{{{\rm{H}}_2}{\rm{O}}}})$, where $x'_{{\rm{H}}_2}$, $x'_{{\rm{H}}_2\text{O}}$ respectively represent the true mole fraction of each species in the gas phase ($N_\text{2}$ might be in the gas phase). The mole fraction is calculated by the mass fraction:
\begin{equation}\label{e2-1-1}
{x'_j} = {Y_j}{M_{\rm{g}}}/{M_j},
\end{equation}
\begin{equation}\label{e2-1-2}
{M_{\rm{g}}} = {\left( {\sum {{Y_j}/{M_j}} } \right)^{ - 1}},
\end{equation}
where $M_\text{g}$ is the average molar mass of the gas mixture, and $Y_j$ is the mass fraction of the gas $j$. $x_\text{eq}$ is the mole fraction of reducing gas at reaction equilibrium, and its relationship with equilibrium constant $K_\text{eq}$ is as follows:
\begin{equation}\label{e2-1-3}
x_{{\rm{eq}},j}^{kl} = 1/(K_{{\rm{eq}}}^{kl} + 1).
\end{equation}
For the three-layer USCM, according to the reaction resistance network in Figure \ref{fig2} (a), the resistance coefficients $v_1^{{kl}}$, $v_2^{{kl}}$ and $v_3^{{kl}}$ can be derived as follows:
\begin{equation}\label{e2-2}
v_1^{{\rm{hm}}} = \left[ {{A_{{\rm{wFe}}}}\left( {{A_{{\rm{mw}}}} + {B_{\rm{w}}} + {B_{{\rm{Fe}}}} + F} \right) + \left( {{B_{{\rm{Fe}}}} + F} \right)\left( {{A_{{\rm{mw}}}} + {B_{\rm{w}}}} \right)} \right]/W,
\end{equation}
\begin{equation}\label{e2-3}
v_2^{{\rm{hm}}} =  - \left[ {{A_{{\rm{wFe}}}}\left( {{B_{\rm{w}}} + {B_{{\rm{Fe}}}} + F} \right) + {B_{\rm{w}}}\left( {{B_{{\rm{Fe}}}} + F} \right)} \right]/W,
\end{equation}
\begin{equation}\label{e2-4}
v_3^{{\rm{hm}}} =  - {A_{{\rm{mw}}}}\left( {{B_{{\rm{Fe}}}} + F} \right)/W,
\end{equation}
\begin{equation}\label{e2-5}
v_1^{{\rm{mw}}} =  - \left[ {{B_{\rm{w}}}\left( {{A_{{\rm{wFe}}}} + {B_{{\rm{Fe}}}} + F} \right) + {A_{{\rm{wFe}}}}\left( {{B_{{\rm{Fe}}}} + F} \right)} \right]/W,
\end{equation}
\begin{equation}\label{e2-6}
v_2^{{\rm{mw}}} = \left[ {\left( {{A_{{\rm{hm}}}} + {B_{\rm{m}}} + {B_{\rm{w}}}} \right)\left( {{A_{{\rm{wFe}}}} + {B_{{\rm{Fe}}}} + F} \right) + {A_{{\rm{wFe}}}}\left( {{B_{{\rm{Fe}}}} + F} \right)} \right]/W,
\end{equation}
\begin{equation}\label{e2-7}
v_3^{{\rm{mw}}} =  - \left( {{A_{{\rm{hm}}}} + {B_{\rm{m}}}} \right)\left( {{B_{{\rm{Fe}}}} + F} \right)/W,
\end{equation}
\begin{equation}\label{e2-8}
v_1^{{\rm{wFe}}} =  - {A_{{\rm{mw}}}}\left( {{B_{{\rm{Fe}}}} + F} \right)/W,
\end{equation}
\begin{equation}\label{e2-9}
v_2^{{\rm{wFe}}} =  - \left( {{A_{{\rm{hm}}}} + {B_{\rm{m}}}} \right)\left( {{B_{{\rm{Fe}}}} + F} \right)/W,
\end{equation}
\begin{equation}\label{e2-10}
v_3^{{\rm{wFe}}} = \left[ {\left( {{A_{{\rm{hm}}}} + {B_{\rm{m}}}} \right)\left( {{A_{{\rm{mw}}}} + {B_{\rm{w}}} + {B_{{\rm{Fe}}}} + F} \right) + {A_{{\rm{mw}}}}\left( {{B_{\rm{w}}} + {B_{{\rm{Fe}}}} + F} \right)} \right]/W,
\end{equation}
where
\begin{equation}\label{e2-11}
W = {A_{{\rm{wFe}}}}\left( {{A_{{\rm{mw}}}} + {B_{\rm{w}}} + {B_{{\rm{Fe}}}} + F} \right) + \left( {{A_{{\rm{mw}}}} + {B_{\rm{w}}}} \right)\left( {{B_{{\rm{Fe}}}} + F} \right).
\end{equation}
Thus,  the total reaction rate of the gas can be expressed as ${\dot m_j} = \dot m_j^{{\rm{hm}}} + \dot m_j^{{\rm{mw}}} + \dot m_j^{{\rm{wFe}}}$.

If the molar fraction of the reducing gas is lower than the equilibrium molar fraction of the gas in reaction w$\rightarrow$Fe at a certain temperature, there may be six resistance networks as shown in Figure \ref{fig2} (b$\sim$g) during the reaction. The reaction rate of reducing gas can be expressed as:
\begin{equation}\label{eb1}
{\dot m_j} = \dot m_j^{{\rm{hm}}} + \dot m_j^{{\rm{mw}}} = \frac{{{\eta_j}{A_{pi}}}}{{{T_{pi}}}}\frac{{{M_j}{p_{\rm{g}}}}}{{{R_{\rm{g}}}}}{\left[ {{v_1}({x_j} - x_{{\rm{eq}},j}^{{\rm{hm}}}) + {v_2}({x_j} - x_{{\rm{eq}},j}^{{\rm{mw}}})} \right]_j}.
\end{equation}
(i) When the molar fraction of the reducing gas is lower than the equilibrium molar fraction of the gas in reaction m$\rightarrow$w, only the reaction h$\rightarrow$m occurs, and the product layer is only m (Figure \ref{fig2} (b)). Thus, ${v_1} = 1/\left( {{A_{{\rm{hm}}}} + {B_{\rm{m}}} + F} \right)$ and ${v_2} = 0$.
(ii) When the molar fraction of the reducing gas is lower than the equilibrium molar fraction of the gas in reaction m$\rightarrow$w, only the reaction h$\rightarrow$m occurs, and the product layers contain m and w (Figure \ref{fig2} (c)). Thus, ${v_1} = 1/\left( {{A_{{\rm{hm}}}} + {B_{\rm{m}}} + {B_{\rm{w}}} + F} \right)$ and ${v_2} = 0$.
(iii) When the molar fraction of the reducing gas is lower than the equilibrium molar fraction of the gas in reaction m$\rightarrow$w, only the reaction h$\rightarrow$m occurs, and the product layer contains m and Fe (Figure \ref{fig2} (d)). Thus, ${v_1} = 1/\left( {{A_{{\rm{hm}}}} + {B_{\rm{m}}} + {B_{\rm{Fe}}} + F} \right)$ and ${v_2} = 0$.
(iv) When the molar fraction of the reducing gas is lower than the equilibrium molar fraction of the gas in reaction m$\rightarrow$w, only the reaction h$\rightarrow$m occurs, and the product layer consists of m, w and Fe (Figure \ref{fig2} (e)). Thus, ${v_1} = 1/\left( {{A_{{\rm{hm}}}} + {B_{\rm{m}}} + {B_{\rm{w}}} + {B_{{\rm{Fe}}}} + F} \right)$ and ${v_2} = 0$.
(v) When the molar fraction of the reducing gas is higher than the equilibrium molar fraction of the gas in reaction m$\rightarrow$w, but lower than that in reaction w$\rightarrow$Fe, both the reaction h$\rightarrow$m and m$\rightarrow$w occur, and the product layer consists of m and w (Figure \ref{fig2} (f)). Thus, ${v_1} = {A_{{\rm{mw}}}}/W$ and ${v_2} = \left( {{A_{{\rm{hm}}}} + {B_{\rm{m}}}} \right)/W$, where
\begin{equation}\label{eb2}
W = {A_{{\rm{mw}}}}\left( {{A_{{\rm{hm}}}} + {B_{\rm{m}}} + {B_{\rm{w}}} + F} \right) + \left( {{A_{{\rm{hm}}}} + {B_{\rm{m}}}} \right)\left( {{B_{\rm{w}}} + F} \right).
\end{equation}
(vi) When the molar fraction of the reducing gas is higher than the equilibrium molar fraction of the gas in reaction m$\rightarrow$w, but lower than that in reaction w$\rightarrow$Fe, both the reaction h$\rightarrow$m and m$\rightarrow$w occur, and the product layer consists of m, w and Fe (Figure \ref{fig2} (g)). Thus, ${v_1} = {A_{{\rm{mw}}}}/W$ and ${v_2} = \left( {{A_{{\rm{hm}}}} + {B_{\rm{m}}}} \right)/W$, where
\begin{equation}\label{eb3}
W = {A_{{\rm{mw}}}}\left( {{A_{{\rm{hm}}}} + {B_{\rm{m}}} + {B_{\rm{w}}} + {B_{{\rm{Fe}}}} + F} \right) + \left( {{A_{{\rm{hm}}}} + {B_{\rm{m}}}} \right)\left( {{B_{\rm{w}}} + {B_{{\rm{Fe}}}} + F} \right).
\end{equation}

The resistance at the reaction interface is:
\begin{equation}\label{e2-12}
{A_{kl,j}} = {\left[ {\frac{1}{{{k_{{\rm{r,}}kl}}}}{{\left( {1 - {f_k}} \right)}^{ - \frac{2}{3}}}\frac{{{K_{{\rm{eq}}}}_{,kl}}}{{1 + {K_{{\rm{eq}}}}_{,kl}}}} \right]_j},
\end{equation}
\begin{equation}\label{e2-13}
{k_{\mathop{\rm r}\nolimits} } = {k_0}\exp ( - {E_{\rm{a}}}/({R_{\rm{g}}}{T_{\rm{p}}})),
\end{equation}
where $k_\text{r}$ and $k_\text{0}$ is the reaction rate constant and pre-exponential factor respectively, and $E_\text{a}$ is the activation energy. Kinetic parameters of each reaction are shown in Table \ref{tab2}.

\begin{table}[]
\centering
\caption{Hematite reduction by steps.}\label{tab2}
\begin{tabular}{cccc}
\hline
Reaction         & $k_\text{0}$ \citep{takenaka1986mathematical}    & $E_\text{a}$ \citep{takenaka1986mathematical}   &$K_\text{eq}$ \citep{zhang2013thermodynamic}      \\ \hline
R1       &29.17      & 66974    &exp($1433.4/T_\text{p}+9.08$)           \\
R2       &15.56      & 75345    &exp($-7393.9/T_\text{p}+7.56$)           \\
R3       &2858      & 117204    &exp($-2023.8/T_\text{p}+1.24$)           \\ \hline
\end{tabular}
\end{table}

Diffusion resistance of reducing gas $j$ through product layer $l$ is:
\begin{equation}\label{e2-14}
{B_{l,j}} = {\left\{ {\left[ {{{\left( {1 - {f_{l - 1}}} \right)}^{ - 1/3}} - {{\left( {1 - {f_l}} \right)}^{ - 1/3}}} \right]\frac{{{r_{\rm{p}}}}}{{D_{{\rm{eff}}}^l}}}, \right\}_j},
\end{equation}
where ${D_{{\rm{eff}}}^l}$ is the effective diffusion coefficient of gas through the product layer $l$. $f_l$ is the fractional reduction degree, which can be calculated by the following equation \citep{valipour2009mathematical}:
\begin{equation}\label{e2-15}
{f_{\rm{h}}} = \frac{{\frac{{3{m_{\rm{m}}}}}{{{M_{\rm{m}}}}} + \frac{{{m_{\rm{W}}}}}{{{M_{\rm{W}}}}} + \frac{{{m_{{\rm{Fe}}}}}}{{{M_{{\rm{Fe}}}}}}}}{{\frac{{2{m_{\rm{h}}}}}{{{M_{\rm{h}}}}} + \frac{{3{m_{\rm{m}}}}}{{{M_{\rm{m}}}}} + \frac{{{m_{\rm{W}}}}}{{{M_{\rm{W}}}}} + \frac{{{m_{{\rm{Fe}}}}}}{{{M_{{\rm{Fe}}}}}}}},
\end{equation}
\begin{equation}\label{e2-16}
{f_{\rm{m}}} = \frac{{\frac{{{m_{\rm{W}}}}}{{{M_{\rm{W}}}}} + \frac{{{m_{{\rm{Fe}}}}}}{{{M_{{\rm{Fe}}}}}}}}{{\frac{{2{m_{\rm{h}}}}}{{{M_{\rm{h}}}}} + \frac{{3{m_{\rm{m}}}}}{{{M_{\rm{m}}}}} + \frac{{{m_{\rm{W}}}}}{{{M_{\rm{W}}}}} + \frac{{{m_{{\rm{Fe}}}}}}{{{M_{{\rm{Fe}}}}}}}},
\end{equation}
\begin{equation}\label{e2-17}
{f_{\rm{w}}}=\frac{{\frac{{{m_{{\rm{Fe}}}}}}{{{M_{{\rm{Fe}}}}}}}}{{\frac{{2{m_{\rm{h}}}}}{{{M_{\rm{h}}}}} + \frac{{3{m_{\rm{m}}}}}{{{M_{\rm{m}}}}} + \frac{{{m_{\rm{W}}}}}{{{M_{\rm{W}}}}} + \frac{{{m_{{\rm{Fe}}}}}}{{{M_{{\rm{Fe}}}}}}}}.
\end{equation}
The overall reduction degree is:
\begin{equation}\label{e2-18}
{f_{{\rm{ov}}}} = \frac{1}{9}{f_{\rm{h}}} + \frac{2}{9}{f_{\rm{m}}} + \frac{6}{9}{f_{\rm{w}}}.
\end{equation}
If magnetite is used as raw iron ore, the above equation becomes
\begin{equation}\label{e2-19}
{f_{{\rm{ov}}}} = \frac{1}{4}{f_{\rm{m}}} + \frac{3}{4}{f_{\rm{w}}}.
\end{equation}
Resistance of reducing agent through gas film is
\begin{equation}\label{e2-20}
{F_j} = {\left( {\frac{1}{{{\beta_{\rm{f}}}}}} \right)_j},
\end{equation}
where $\beta_\text{f}$ is the gas film mass transfer coefficient, which can be calculated as:
\begin{equation}\label{e2-21}
{\beta _{\rm{f}}} = \frac{{Sh{D_{{\rm{eff}},{\rm{f}}}}}}{{{d_{\rm{p}}}}},
\end{equation}
where the Sherwood number is \citep{Gunn1978}:
\begin{equation}\label{e2-22}
Sh = \left( {7 - 10{\varepsilon _{\rm{g}}} + 5\varepsilon _{\rm{g}}^2} \right)\left( {1 + 0.7R{e_{\rm{p}}}^{0.2}S{c^{1/3}}} \right) + \left( {1.33 - 2.4{\varepsilon _{\rm{g}}} + 1.2\varepsilon _{\rm{g}}^2} \right)R{e_{\rm{p}}}^{0.7}S{c^{1/3}}.
\end{equation}
Particle Reynolds number and Schmidt number are respectively
\begin{equation}\label{e2-23}
R{e_{\rm{p}}} = \frac{{{\rho _{\rm{g}}}{\varepsilon _{\rm{g}}}{d_{\rm{p}}}\left| {{{\bf{u}}_{\rm{g}}} - {{\bf{v}}_{\rm{p}}}} \right|}}{{{\mu _{\rm{g}}}}},
\end{equation}
\begin{equation}\label{e2-24}
Sc = \frac{{{\mu _{\rm{g}}}}}{{{\rho _{\rm{g}}}{D_{{\rm{eff}},{\rm{f}}}}}},
\end{equation}
where $D_\text{eff,f}$ is the effective diffusion coefficient of reducing agent through the gas film, which is calculated by Equation \ref{e2-27}. The mass change rate of the other component $o$ in reaction $k$$\rightarrow$$l$ is expressed by the mass change rate of the reducing gas:
\begin{equation}\label{e2-25}
\dot m_o^{kl} = \frac{{\eta _o^{kl}{M_o}}}{{\eta _j^{kl}{M_j}}}\dot m_j^{kl}.
\end{equation}

\subsection{Reaction rate model: the raw iron ore is magnetite}
\begin{figure}
\centerline{\includegraphics[width=0.6\textwidth]{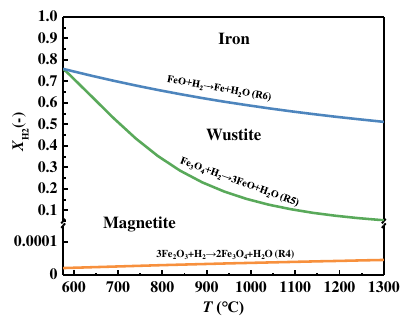}}
\caption{Thermodynamic equilibrium phase diagram of hematite reduced by $\rm{H_2} $, with the equilibrium constant cited from the fitting data of \cite{zhang2013thermodynamic} ($T>570^\circ \text{C}$ ).}\label{fig3}
\end{figure}
\begin{figure}
\centerline{\includegraphics[width=0.6\textwidth]{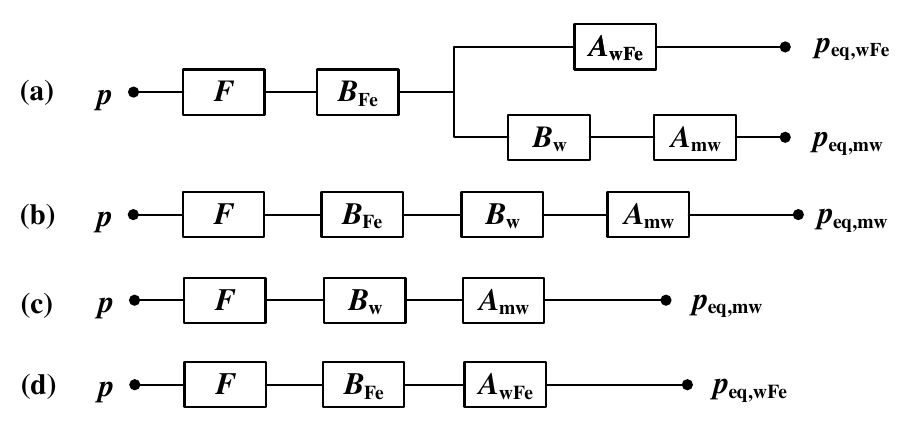}}
\caption{Resistance network for progressive reduction of magnetite.}\label{fig4}
\end{figure}
According to the thermodynamic equilibrium phase diagram of hematite reduced by $\rm{H_2}$ (Figure \ref{fig3}), if the raw iron ore consists of $\rm{Fe_3O_4}$ and FeO, the reaction resistance network is shown as Figure \ref{fig4}. The reaction rate of reducing gas can be expressed as:
\begin{equation}\label{ea1}
{\dot m_j} = \dot m_j^{{\rm{mw}}} + \dot m_j^{{\rm{wFe}}} = \frac{{{\eta _j}{A_{pi}}}}{{{T_{pi}}}}\frac{{{M_j}{p_{\rm{g}}}}}{{{R_{\rm{g}}}}}{\left[ {{v_2}({x_j} - x_{{\rm{eq}},j}^{{\rm{mw}}}) + {v_3}({x_j} - x_{{\rm{eq}},j}^{{\rm{wFe}}})} \right]_j}.
\end{equation}
(i) When the molar fraction of the reducing gas is higher than the equilibrium molar fraction of the gas in reaction w$\rightarrow$Fe, both the reaction m$\rightarrow$w and w$\rightarrow$Fe occur (Figure \ref{fig4} (a)). Thus, ${v_2} = {A_{{\rm{wFe}}}}/W$, and ${v_3} = \left( {{A_{{\rm{mw}}}} + {B_{\rm{w}}}} \right)/W$ in above equation, where
\begin{equation}\label{ea2}
W = {A_{{\rm{wFe}}}}\left( {{A_{{\rm{mw}}}} + {B_{\rm{w}}} + {B_{{\rm{Fe}}}} + F} \right) + \left( {{A_{{\rm{mw}}}} + {B_{\rm{w}}}} \right)\left( {{B_{{\rm{Fe}}}} + F} \right).
\end{equation}
(ii) When the molar fraction of the reducing gas is lower than the equilibrium molar fraction of the gas in reaction w$\rightarrow$Fe, only the reaction m$\rightarrow$w occurs (Figure \ref{fig4} (b)). Thus, ${v_2} = 1/\left( {{A_{{\rm{mw}}}} + {B_{\rm{w}}} + {B_{{\rm{Fe}}}} + F} \right)$ and $v_3$ = 0.
(iii) When the molar fraction of the reducing gas is lower than the equilibrium molar fraction of the gas in reaction w$\rightarrow$Fe, only the reaction m$\rightarrow$w occurs, and the product layer does not contain Fe (Figure \ref{fig4} (c)). Thus, ${v_2} = 1/\left( {{A_{{\rm{mw}}}} + {B_{\rm{w}}} + F} \right)$ and $v_3$ = 0.
(iv) When m has been completely converted to w, only the reaction m$\rightarrow$w occurs (Figure \ref{fig4} (d)). Thus, $v_2$ = 0, and ${v_3} = 1/({A_{{\rm{wFe}}}} + {B_{{\rm{Fe}}}} + F)$.

\subsection{Gas diffusion in a porous particle}\label{s2-2}
The diffusion of gas through porous solids depends on pore structure, porosity, tortuosity and pore size distribution. Gas diffusion within pores can be divided into two forms  according to the relative magnitude of pore diameter ($d_\text{a}$) and the mean free path of molecular motion ($\lambda$). When $\lambda_j/d_\text{a}\leq$0.01 ($\lambda_j=1.013/(p_\text{g}x_j)$, cm), molecular diffusion is dominant \citep{li2017reaction}, and the Fuller-Schettler-Giddings correlation \citep{fuller1966new} is used to calculate the bimolecular diffusion coefficient (unit: $\rm{m}^2$/s):
\begin{equation}\label{e2-26}
{D_{ij}} = \frac{{{{10}^{{\rm{ - }}7}}T_{\rm{g}}^{1.75}{{\left( {M_i^{ - 1} + M_j^{ - 1}} \right)}^{1/2}}}}{{\left( {{p_{\rm{g}}}/101325} \right){{\left( {V_{{\rm{d,}}i}^{1/3} + V_{{\rm{d}},j}^{1/3}} \right)}^2}}},
\end{equation}
where $M_i$ and $M_j$ are expressed in g/mol, and $V_{\text{d,}i}$ and $V_{\text{d,}j}$ ($\rm{cm}^3$/mol) are diffusion volume of gas $i$ and $j$, respectively. In this study, ${V_{{\rm{d,}}{{\rm{H}}_{\rm{2}}}}}$=7.07, ${V_{{\rm{d,}}{{\rm{H}}_{\rm{2}}\rm{O}}}}$=12.7 and ${V_{{\rm{d,}}{{\rm{N}}_{\rm{2}}}}}$=17.9. For a multi-component gas mixture, the molecular diffusion coefficient is \citep{wilke1955correlation}:
\begin{equation}\label{e2-27}
D_i^{\rm{m}} = (1 - {x_i}){\left( {\sum\limits_{i \ne j} {{x_j}/{D_{ij}}} } \right)^{ - 1}}.
\end{equation}
While $\lambda_j/d_\text{a}\geq$10, diffusion inside the pore is called the Knudsen diffusion \citep{li2017reaction}. Most collisions occur between molecules and pore wall, and the collisions between molecules have little impact on the transport process. Therefore the Knudsen diffusion coefficient is determined by the pore radius, and not influenced by other gas species in the system. Its value can be calculated by
\begin{equation}\label{e2-28}
D_i^{\rm{K}} = 0.97{r_{\rm{a}}}\sqrt {{T_{\rm{g}}}/{M_i}},
\end{equation}
where $r_\text{a}$ is pore radius (unit: cm). If the relative magnitude of $\lambda$ and $d_\text{a}$ lies between these two extreme cases, both molecular and Knudsen diffusions should be taken into account, and the total diffusion coefficient for species $i$ is:
\begin{equation}\label{e2-29}
{D_i} = \frac{1}{{1/D_i^{\rm{m}} + 1/D_i^{\rm{K}}}},
\end{equation}
where the effective diffusion coefficient of gas $i$ through product layer $l$ is \citep{wakao1962diffusion}
\begin{equation}\label{e2-30}
D_{{\rm{eff,}}i}^l = {D_i}{\xi _l}/\tau.
\end{equation}
where $\tau$ is tortuosity. $\xi_l$ is the porosity of $l$, its value varies with time and can be calculated based on the initial porosity of the particle. For porous particle, there is
\begin{equation}\label{e2-31}
{\xi _l} = 1 - {\rho _{{\rm{eff,}}l}}/{\rho _{{\rm{true,}}l}},
\end{equation}
where $\rho_{\text{eff},l}$ and $\rho_{\text{true},l}$ are the effective and true density of particle layer $l$, respectively. Assuming that the porosity of hematite core is equal to the porosity of the particle, the effective density of hematite core is:
\begin{equation}\label{e2-32}
{\rho _{{\rm{eff,h}}}} = (1 - {\xi _{\rm{p}}}){\rho _{{\rm{true,h}}}}.
\end{equation}
The effective density of each product layer is respectively \citep{kinaci2020cfd}
\begin{equation}\label{e2-33}
{\rho _{{\rm{eff,m}}}} = {\rho _{{\rm{eff,h}}}}{q_{{\rm{m,h}}}},
\end{equation}
\begin{equation}\label{e2-34}
{\rho _{{\rm{eff,w}}}} = {\rho _{{\rm{eff,m}}}}{q_{{\rm{w,m}}}},
\end{equation}
\begin{equation}\label{e2-35}
{\rho _{{\rm{eff,Fe}}}} = {\rho _{{\rm{eff,w}}}}{q_{{\rm{Fe,w}}}},
\end{equation}
where $q_\text{b,a}$ represents the equivalent mass of product b when the consumption of reactant a is 1, which can be calculated by \citep{kinaci2020cfd}
\begin{equation}\label{e2-36}
{q_{b,a}}{\rm{ = }}\frac{{\left| {{\eta _b}} \right|{M_b}}}{{\left| {{\eta _a}} \right|{M_a}}}.
\end{equation}
Accordingly, it is easy to calculate the values of $q_\text{m,h}$, $q_\text{w,m}$ and $q_\text{Fe,w}$ as 0.9666, 0.9309 and 0.7773, respectively, and thus the porosity of each product layer was obtained by combining Equations \ref{e2-31}, \ref{e2-33}$\sim$\ref{e2-35}.

\section{CFD-DEM method}\label{s3}

\subsection{Governing equations for particles}\label{s3-1}
According to the conservation of particle mass, there are:
\begin{equation}\label{e3-1}
{\dot m_{\rm{p}}} = \sum\limits_z {{{\dot m}_z}},
\end{equation}
\begin{equation}\label{e3-2}
{\dot m_z}{\rm{ = }}\sum {\dot m_z^{kl}},
\end{equation}
where $\dot m_{\rm{p}}$ is the mass change rate of the particle, $z$ represents the particle component $\rm{Fe_2O_3}$, $\rm{Fe_3O_4}$, FeO or Fe, and $\dot m_z^{kl}$ is the mass change rate of $z$ in the reaction $k$$\rightarrow$$l$, calculated by Equation \ref{e2-25}.

The translational and rotational motion of particles can be described by the Newton's second law. The translational equation is written as:
\begin{equation}\label{e3-3}
\frac{{d({m_{\rm{p}}}{{\bf{v}}_{\rm{p}}})}}{{dt}} = {{\bf{F}}_{\rm{g}}} + {{\bf{F}}_p} + \sum\nolimits_{b{\rm{ = 1}}}^{{n_b}} {{{\bf{F}}_{{\rm{c,ab}}}}}  + {{\bf{F}}_{{\rm{d,p}}}}
\end{equation}

\begin{equation}\label{e3-4}
{m_{\rm{p}}} = {m_{\rm{0}}} + {\dot m_{\rm{p}}}\Delta {t_{{\rm{CFD}}}},
\end{equation}
where $m_\text{p}$ and $\bf{v}_\text{p}$ are the particle mass and velocity, respectively. $m_\text{0}$ is the particle mass of the previous CFD step, ${\Delta t}_{\text{CFD}}$ is the time step of gas phase evolution. The gravity ${{\bf{F}}_{\rm{g}}} = {m_{\rm{p}}}{\bf{g}}$, and the pressure gradient force ${{\bf{F}}_p}=-{V_{\rm{p}}}\nabla {p_{\rm{g}}}$. The contact force between particle $a$ and $b$ is expressed as ${{\bf{F}}_{{\rm{c,ab}}}}{\rm{ = }}{{\bf{F}}_{{\rm{n,ab}}}} + {{\bf{F}}_{{\rm{t,ab}}}}$, where the details of normal contact force (${\bf{F}}_{{\rm{n,ab}}}$) and tangential contact force (${\bf{F}}_{{\rm{t,ab}}}$) can be seen in \cite{lan2020long}. The gas-solid drag force is:
\begin{equation}\label{e3-5}
{{\bf{F}}_{{\rm{d,p}}}} = \frac{{\beta {V_{\rm{p}}}}}{{1 - {\varepsilon _{\rm{g}}}}}({{\bf{u}}_{\rm{g}}} - {{\bf{v}}_{\rm{p}}})
\end{equation}
where the drag coefficient $\beta$ is calculated by Gidaspow's correlation \cite{gidaspow1994multiphase}, i.e.
\begin{equation}\label{e3-5-1}
\beta  = \left\{ \begin{array}{l}
150\frac{{\varepsilon _{\rm{p}}^2{\mu _{\rm{g}}}}}{{{\varepsilon _{\rm{g}}}d_{\rm{p}}^2}} + 1.75\frac{{{\varepsilon _{\rm{g}}}{\rho _{\rm{g}}}\left| {{{\bf{u}}_{\rm{g}}} - {{\bf{v}}_{\rm{p}}}} \right|}}{{{d_{\rm{p}}}}},{\varepsilon _{\rm{g}}} < 0.8\\
\frac{3}{4}{C_{\rm{D}}}\frac{{{\varepsilon _{\rm{p}}}{\varepsilon _{\rm{g}}}{\rho _{\rm{g}}}\left| {{{\bf{u}}_{\rm{g}}} - {{\bf{v}}_{\rm{p}}}} \right|}}{{{d_{\rm{p}}}}}\varepsilon _{\rm{g}}^{ - 2.65},{\varepsilon _{\rm{g}}} > 0.8,
\end{array} \right.
\end{equation}
If the iron ore is non-spherical, the Gidaspow model is modified with the sphericity to obtain the Gidaspow-Ganser drag force correlation \citep{Ganser1993A,Hua2015Eulerian}, that is:
\begin{equation}\label{e3-5-2}
\beta  = \left\{ \begin{array}{l}
150\frac{{\varepsilon _{\rm{p}}^2{\mu _{\rm{g}}}}}{{{\varepsilon _{\rm{g}}}d_{\rm{p}}^2{\psi ^2}}} + 1.75\frac{{{\varepsilon _{\rm{g}}}{\rho _{\rm{g}}}\left| {{{\bf{u}}_{\rm{g}}} - {{\bf{v}}_{\rm{p}}}} \right|}}{{{d_{\rm{p}}}\psi }},{\varepsilon _{\rm{g}}} < 0.8\\
\frac{3}{4}{C_{\rm{D}}}\frac{{{\varepsilon _{\rm{p}}}{\varepsilon _{\rm{g}}}{\rho _{\rm{g}}}\left| {{{\bf{u}}_{\rm{g}}} - {{\bf{v}}_{\rm{p}}}} \right|}}{{{d_{\rm{p}}}}}\varepsilon _{\rm{g}}^{ - 2.65},{\varepsilon _{\rm{g}}} > 0.8,
\end{array} \right.
\end{equation}
where $\mit{\Psi}$ is the sphericity of the particle, and the drag coefficient of a single particle $C_\text{D}$ is calculated by Ganser correlation \citep{Ganser1993A}:
\begin{equation}\label{e3-5-3}
{C_{\rm{D}}} = \frac{{24}}{{R{e_{\rm{p}}}{K_1}}}\left[ {1.0 + 0.1118{{(R{e_{\rm{p}}}{K_1}{K_2})}^{0.6567}}} \right] + \frac{{0.4305{K_2}}}{{1 + 3305/(R{e_{\rm{p}}}{K_1}{K_2})}},
\end{equation}
\begin{equation}\label{e3-5-4}
{K_1} = {\left( {\frac{1}{3} + \frac{2}{3}{\psi ^{ - 0.5}}} \right)^{ - 1}} - 2.25\frac{{{d_{\rm{p}}}}}{D},
\end{equation}
\begin{equation}\label{e3-5-5}
{K_2} = {10^{1.8148{{( - \log \psi )}^{0.5743}}}},
\end{equation}
where $D$ is the equivalent-area bed diameter. In addition, the momentum exchange caused by mass exchange is 0, so this term does not appear on the right side of Equation \ref{e3-3} according to the discussion of \cite{lan2022cfd}.

The particle rotation equation is \citep{xu2019virtual}
\begin{equation}\label{e3-6}
\frac{{d({{\boldsymbol{\omega }}_{\rm{p}}}{I_{\rm{p}}})}}{{dt}} = {{\bf{R}}_{\rm{p}}} \times {{\bf{F}}_{\rm{t}}} - {\mu _{\rm{r}}}{r_{\rm{p}}}\left| {{{\bf{F}}_{\rm{n}}}} \right|{{\boldsymbol {\omega }}_{\rm{p}}},
\end{equation}
where
\begin{equation}\label{e3-7}
{I_{\rm{p}}} = \frac{2}{5}{m_{\rm{p}}}r_{\rm{p}}^2,
\end{equation}
where $\boldsymbol {\omega}_\text{p}$  and $I_\text{p}$ are the angular velocity and moment of inertia of the particle, respectively.

The particle energy conservation equation is
\begin{equation}\label{e3-8}
\frac{{d({m_{\rm{p}}}{C_{p,{\rm{p}}}}{T_{\rm{p}}})}}{{dt}} = {Q_{{\rm{conv}}}} + \sum\nolimits_{b{\rm{ = 1}}}^{{n_b}} {{Q_{ab}}}  + {Q_{\rm{r}}} + {Q_{{\rm{rm}}}} + {Q_{{\rm{pw}}}} + {Q_{{\rm{rad}}}},
\end{equation}
\begin{equation}\label{e3-9}
{C_{p,{\rm{p}}}} = \sum {{Y_z}{C_{p,z}}},
\end{equation}
where $Y_z$ and $C_{p,z}$ are the mass fraction and heat capacity of particle component $z$ respectively, and $C_{p,z}$ is a function of particle temperature, which can be calculated by Equation \ref{e3-23} and Table \ref{tab c1}. The rate of heat flow due to gas convection is:
\begin{equation}\label{e3-10}
{Q_{{\rm{conv}}}} = {h_{\rm{p}}}{A_{\rm{p}}}({T_{\rm{g}}} - {T_{\rm{p}}}),
\end{equation}
with
\begin{equation}\label{e3-11}
{h_{\rm{p}}} = \frac{{Nu{\lambda _{\rm{g}}}}}{{{d_{\rm{p}}}}},
\end{equation}
where $\lambda_\text{g}$ is the thermal conductivity coefficient of the gas phase, and $Nu$ can be calculated by Gunn's empirical correlation \citep{Gunn1978}:
\begin{equation}\label{e3-12}
Nu = \left( {7 - 10{\varepsilon _{\rm{g}}} + 5\varepsilon _{\rm{g}}^2} \right)\left( {1 + 0.7R{e_{\rm{p}}}^{0.2}P{r^{1/3}}} \right) + \left( {1.33 - 2.4{\varepsilon _{\rm{g}}} + 1.2\varepsilon _{\rm{g}}^2} \right)R{e_{\rm{p}}}^{0.7}P{r^{1/3}},
\end{equation}
with
\begin{equation}\label{e3-13}
Pr = \frac{{{C_{p,{\rm{g}}}}{\mu _{\rm{g}}}}}{{{\lambda _{\rm{g}}}}}.
\end{equation}
The contact heat transfer rate between the particle $a$ and $b$ is the sum of heat conduction rates of particle-particle and particle-fluid-particle, i.e. ${Q_{ab}} = Q_{ab}^{{\rm{pp}}} + Q_{ab}^{{\rm{pfp}}}$, where the particle-particle heat transfer rate is \citep{batchelor1977thermal}:
\begin{equation}\label{e3-14}
Q_{ab}^{{\rm{pp}}} = 4{R_{{\rm{c,}}ab}}{\lambda _{{\rm{e}},ab}}({T_b} - {T_a}),
\end{equation}
and the particle-fluid-particle heat transfer rate is \citep{rong1999simulation}:
\begin{equation}\label{e3-15}
Q_{ab}^{{\rm{pfp}}} = 2\pi {\lambda _{\rm{g}}}\left( {{T_b} - {T_a}} \right) \times \int_{{R_{{\rm{in}}}}}^{{R_{{\rm{out}}}}} {\frac{R}{{{d_{ab}} - \sqrt {R_a^2 - {R^2}}  - \sqrt {R_b^2 - {R^2}} }}dR}.
\end{equation}
The heat conduction rate between particle $a$ and the wall is the sum of heat conduction rates of the particle-wall and particle-fluid-wall, that is, ${Q_{a{\rm{w}}}} = Q_{a{\rm{w}}}^{{\rm{pw}}} + Q_{a{\rm{w}}}^{{\rm{pfw}}}$, where \citep{batchelor1977thermal,rong1999simulation}
\begin{equation}\label{e3-15-1}
Q_{a{\rm{w}}}^{{\rm{pw}}} = 4{R_{{\rm{c,}}a{\rm{w}}}}{\lambda _{{\rm{e}},a{\rm{w}}}}({T_{\rm{w}}} - {T_a}),
\end{equation}
\begin{equation}\label{e3-15-2}
Q_{a{\rm{w}}}^{{\rm{pfw}}}{\rm{ = }}2\pi {\lambda _{\rm{g}}}\left( {{T_{\rm{w}}} - {T_a}} \right)\int_{{r_{{\rm{in,w}}}}}^{{r_{{\rm{out,w}}}}} {\frac{{{r_{\rm{w}}}}}{{{L_{\rm{w}}}}}dr}.
\end{equation}
The physical meaning and calculation of each variable in Equation \ref{e3-14}, \ref{e3-15}, \ref{e3-15-1} and \ref{e3-15-2} can be found in \cite{zhao2020Acomputational}. The radiation heat flow rate of single particle from the environment is:
\begin{equation}\label{e3-16}
{Q_{{\rm{rad}}}} = \sigma {\eta _{\rm{p}}}{A_{\rm{p}}}(T_{{\rm{se,p}}}^4 - T_{\rm{p}}^4),
\end{equation}
where $\sigma$ represents the Stefan-Boltzmann constant ($5.67\times10^{-8}$ W/$\rm{m^2K^4}$), and $\eta_\text{p}$ is the particle emissivity. $T_\text{se,p}$ is the average temperature of particles and fluid in a sphere with a volume of $\Omega$ around the particles, which is calculated as \citep{zhou2009particle}:
\begin{equation}\label{e3-17}
{T_{{\rm{se,p}}}}{\rm{ = }}{\varepsilon _{\rm{g}}}{T_{{\rm{g,}}\Omega }} + (1 - {\varepsilon _{\rm{g}}})\frac{1}{{{n_{{\rm{p,}}\Omega }}}}\sum\limits_{a = 1}^{{n_{{\rm{p,}}\Omega }}} {{T_{{\rm{p,}}a}}},
\end{equation}
where $\varepsilon_\text{g}$, $T_\text{g}$, $\Omega$ and $n_{\rm{p,}\Omega}$ are the gas volume fraction, gas temperature and number of particles in the domain $\Omega$, respectively. The radius of the sphere domain is generally chosen as 1.5$d_\text{p}$ \citep{zhou2009particle}. The heat flow rate caused by reduction reaction is:
\begin{equation}\label{e3-18}
{Q_{\rm{r}}} = \sum {Q_{\rm{r}}^{kl}},
\end{equation}
\begin{equation}\label{e3-19}
Q_{\rm{r}}^{kl} =  - \sum\limits_z {{H_z}\dot m_z^{kl}}.
\end{equation}
The heat transfer rate caused by mass transfer is given by \citep{musser2015constitutive}
\begin{equation}\label{e3-20}
{Q_{{\rm{rm}}}} = \sum {Q_{{\rm{rm}}}^{kl}},
\end{equation}
\begin{equation}\label{e3-21}
Q_{{\rm{rm}}}^{kl} =  - \sum\limits_j {{H_j}\dot m_j^{kl}},
\end{equation}
where the subscripts $z$ and $j$ represent the solid and gas species in the reaction $k$$\rightarrow$$l$, respectively. $H$ is the standard formation enthalpy, which is a polynomial function of temperature and can be written as \citep{burcat2005third}:
\begin{equation}\label{e3-22}
H = {R_{\rm{g}}}({a_1}T + 1/2{a_2}{T^2} + 1/3{a_3}{T^3} + 1/4{a_4}{T^4} + 1/5{a_5}{T^5} + {a_6})/M,
\end{equation}
where $M$ is the molar mass of a species and the relationship between specific heat and temperature is as follows:
\begin{equation}\label{e3-23}
{C_p} = {R_{\rm{g}}}({a_1} + {a_2}T + {a_3}{T^2} + {a_4}{T^3} + {a_5}{T^4})/M.
\end{equation}
The value of $a_i$ is given in Table \ref{tab c1}.

\subsection{Governing equations for gases}\label{s3-2}
In the CFD-DEM method, the gas phase is described by continuum model, and the mass conservation equation expressed by local average variable in a computational grid is:
\begin{equation}\label{e3-24}
\frac{\partial }{{\partial t}}({\varepsilon _{\rm{g}}}{\rho _{\rm{g}}}) + \nabla  \cdot ({\varepsilon _{\rm{g}}}{\rho _{\rm{g}}}{{\bf{u}}_{\rm{g}}}) = \sum {{M_{{\rm{m,}}j}}},
\end{equation}
where
\begin{equation}\label{e3-25}
{M_{{\rm{m,}}j}} = \frac{1}{{{V_{{\rm{cell}}}}}}\sum\nolimits_{pi{\rm{ = 1}}}^{{n_{{\rm{p,cell}}}}} {{{\dot m}_{j,pi}}},
\end{equation}
where $M_{\text{m,}j}$ is the cell-averaged mass transfer rate of reactant or product gas $j$ ($\rm{H_2}$ or $\rm{H_2O}$), and $n_\text{p,cell}$ the number of particles in a gas cell.

The momentum equation is written as:
\begin{equation}\label{e3-26}
\frac{\partial }{{\partial t}}({\varepsilon _{\rm{g}}}{\rho _{\rm{g}}}{{\bf{u}}_{\rm{g}}}) + \nabla  \cdot ({\varepsilon _{\rm{g}}}{\rho _{\rm{g}}}{{\bf{u}}_{\rm{g}}}{{\bf{u}}_{\rm{g}}}) + \nabla  \cdot \left( {{\varepsilon _{\rm{g}}}{\rho _{\rm{g}}}{\bf{u}}_{\rm{g}}^\prime {\bf{u}}_{\rm{g}}^\prime } \right) =  - {\varepsilon _{\rm{g}}}\nabla p + \nabla  \cdot ({\varepsilon _{\rm{g}}}{{\bf{\tau }}_{\rm{g}}}) + {\varepsilon _{\rm{g}}}{\rho _{\rm{g}}}{\bf{g}} + {{\bf{M}}_{\rm{F}}},
\end{equation}
where
\begin{equation}\label{e3-27}
{{\bf{M}}_{\rm{F}}} = {K_{g{\rm{p}}}}({{\bf{u}}_{\rm{g}}} - {{\bf{u}}_{\rm{p}}}),
\end{equation}
\begin{equation}\label{e3-27a}
{K_{g{\rm{p}}}} =  - \frac{1}{{{V_{{\rm{cell}}}}\left| {{{\bf{u}}_{\rm{g}}} - {{\bf{u}}_{\rm{p}}}} \right|}}\sum\limits_{pi{\rm{ = 1}}}^{{n_{{\rm{p,cell}}}}} {\frac{{\beta {V_{pi}}}}{{1 - {\varepsilon _{\rm{g}}}}}\left| {{{\bf{u}}_{\rm{g}}} - {{\bf{v}}_{pi}}} \right|}.
\end{equation}
The third term on the left side of Equation \ref{e3-26} arises due to the pseudo-turbulence, which is usually neglected in CFD-DEM simulations due to the lack of a suitable closure. However, recent PR-DNS (Particle-resolved direct numerical simulation) data show that gas-phase velocity fluctuations can contribute significantly, even in laminar gas-solid flows. Mehrabadi \cite{mehrabadi2015pseudo} provided a closure for the PTRS components using fixed-bed simulations which is also validated as an approximation to high stokes number freely evolving suspensions. ${{\bf{u}}_{\rm{g}}^\prime }$ is the pseudo-turbulence fluctuation of velocity, and the component form of pseudo-turbulent Reynolds stress (PTRS) tensor ${\bf{R}} = {\bf{u}}_{\rm{g}}^\prime {\bf{u}}_{\rm{g}}^\prime$ can be expressed as \citep{mehrabadi2015pseudo}:
\begin{equation}\label{e3-28}
{R_{ij}} = 2{k_{\rm{g}}}\left( {{b_{ij}} + \frac{1}{3}{\delta _{ij}}} \right),
\end{equation}
where $k_\text{g}$ is the gas phase pseudo-turbulent kinetic energy, which is modeled by
\begin{equation}\label{e3-29}
\frac{{{k_{\rm{g}}}}}{{{E_{\rm{g}}}}} = 2{\varepsilon _{\rm{p}}} + 2.5{\varepsilon _{\rm{p}}}\varepsilon _{\rm{g}}^3\exp \left( { - {\varepsilon _{\rm{p}}}Re_{\rm{m}}^{1/2}} \right),
\end{equation}
\begin{equation}\label{e3-30}
{E_{\rm{g}}} = \frac{1}{2}{\left| {{{\bf{u}}_{\rm{g}}} - {{\bf{u}}_{\rm{p}}}} \right|^2},
\end{equation}
where ${{{\bf{u}}_{\rm{p}}}}$ is the cell-averaged particle velocity, and the components of anisotropy tensor parallel to and perpendicular to the mean slip velocity are given by
\begin{equation}\label{e3-31}
{b_{{\rm{pa}}}} = \frac{{0.523}}{{1 + 0.305\exp \left( { - 0.114R{e_{\rm{m}}}} \right)}}\exp \left[ {\frac{{ - 3.511{\varepsilon _{\rm{p}}}}}{{1 + 1.801\exp \left( { - 0.005R{e_{\rm{m}}}} \right)}}} \right],
\end{equation}
\begin{equation}\label{e3-32}
{b_{{\rm{pe}}}} =  - \frac{1}{2}{b_{{\rm{pa}}}},
\end{equation}
where $Re_\text{m}$ is the Reynolds number based on cell-averaged slip velocity and size of particles. The PTRS tensor $\bf{R}$ is the diagonal tensor, whose off-diagonal elements are zero, i.e. $\bf{R}= \rm{diag}(R_\text{pa},R_\text{pe},R_\text{pe})$. It should be noted that the orientation of the axis of the tensor $\bf{R}$ is a local coordinate system with respect to the slip velocity, which needs to be converted to the global Cartesian coordinate system using Gram-Schmidt method \citep{peng2019implementation}. The viscous stress in the gas phase on the right side of the Equation \ref{e3-26} is:
\begin{equation}\label{e3-33}
{{\boldsymbol{\tau }}_{\rm{g}}} = {\mu _{\rm{g}}}(\nabla {{\bf{u}}_{\rm{g}}} + \nabla {\bf{u}}_{\rm{g}}^{\rm{T}}{\rm{) - }}\frac{2}{3}{\mu _{\rm{g}}}(\nabla  \cdot {{\bf{u}}_{\rm{g}}}){\bf{I}},
\end{equation}
where $\bf{I}$ is the unit tensor. $K_\text{gp}$ in Equation \ref{e3-27} is the momentum transfer coefficient, and $\bf{u}_\text{p}$ is the cell-averaged particle velocity. It is believed that the momentum exchange rate caused by the interphase mass transfer is 0, and the specific explanation can refer to a relevant literature \citep{lan2022cfd}.

The energy equation for gas phase is:
\begin{equation}\label{e3-34}
\frac{\partial }{{\partial t}}({\varepsilon _{\rm{g}}}{\rho _{\rm{g}}}{C_{p,{\rm{g}}}}{T_{\rm{g}}}) + \nabla  \cdot ({\varepsilon _{\rm{g}}}{\rho _{\rm{g}}}{{\bf{u}}_{\rm{g}}}{C_{p,{\rm{g}}}}{T_{\rm{g}}}) + \nabla  \cdot ({\varepsilon _{\rm{g}}}{\rho _{\rm{g}}}{C_{p,{\rm{g}}}}{\bf{u}}_{\rm{g}}^\prime T_{\rm{g}}^\prime ) = \nabla  \cdot ({\varepsilon _{\rm{g}}}{\lambda _{\rm{g}}}\nabla {T_{\rm{g}}}) + {M_{\rm{h}}} + {M_{\rm{r}}} + {M_{{\rm{r,m}}}}.
\end{equation}
The third term on the left side of the above equation arises due to pseudo-turbulence, where ${{T}_{\rm{g}}^\prime }$ is the pseudo-turbulence fluctuation of temperature, and ${\bf{R}_T} = {\bf{u}}_{\rm{g}}^\prime {T}_{\rm{g}}^\prime$ is pseudo-turbulence heat flux (PTHF) tensor. Sun \cite{sun2015modeling} used particle-resolved direct numerical simulation (PR-DNS) method to develop a closed correlation suitable for $0<\varepsilon_\text{p}\leq0.5$ and $1\leq Re_\text{m} < 100$, where the PTHF component parallel to the mean slip velocity is written as:
\begin{equation}\label{e3-35}
\begin{array}{l}
\frac{{{\alpha _{{\rm{pa}}}}}}{{{\alpha _{\rm{g}}}}} = \frac{{1.4R{e_{\rm{m}}}\left( {R{e_{\rm{m}}} + 1.4} \right)Pr\exp \left( { - 0.002089R{e_{\rm{m}}}} \right)\left[ {{\varepsilon _{\rm{g}}}\left( { - 5.11{\varepsilon _{\rm{p}}} + 10.1\varepsilon _{\rm{p}}^2 - 10.85\varepsilon _{\rm{p}}^3} \right) + 1 - \exp \left( { - 10.96{\varepsilon _{\rm{p}}}} \right)} \right]}}{{3\pi \varepsilon _{\rm{g}}^2Nu\left( {1.17{\varepsilon _{\rm{p}}} - 0.2021\varepsilon _{\rm{p}}^{1/2} + 0.08568\varepsilon _{\rm{p}}^{1/4}} \right)\left[ {1 - 1.6{\varepsilon _{\rm{p}}}{\varepsilon _{\rm{g}}} - 3{\varepsilon _{\rm{p}}}\varepsilon _{\rm{g}}^4\exp \left( { - Re_{\rm{m}}^{0.4}{\varepsilon _{\rm{p}}}} \right)} \right]}}
\end{array},
\end{equation}
\begin{equation}\label{e3-36}
Nu = \left( { - 0.46 + 1.77{\varepsilon _{\rm{g}}} + 0.69\varepsilon _{\rm{g}}^2} \right)\varepsilon _{\rm{g}}^{{\rm{ - }}3} + \left( {1.37 - 2.4{\varepsilon _{\rm{g}}} + 1.2\varepsilon _{\rm{g}}^2} \right)R{e_{\rm{m}}}^{0.7}P{r^{1/3}}.
\end{equation}
Where $\alpha_\text{g}$ is thermal diffusion coefficient of gas phase ($\alpha_\text{g} = \lambda_\text{g}/(\rho_\text{g}C_{p\text{,g}})$). The PTHF component perpendicular to the mean slip velocity is:
\begin{equation}\label{e3-37}
{\alpha _{{\rm{pe}}}} = \left( {\frac{{{R_{{\rm{pe}}}}}}{{{R_{{\rm{pa}}}}}}} \right){\alpha _{{\rm{pa}}}} = \left( {\frac{{3{b_{{\rm{pe}}}} + 1}}{{3{b_{{\rm{pa}}}} + 1}}} \right){\alpha _{{\rm{pa}}}}.
\end{equation}
The PTHF tensor ${{\bf{R}}_T} = {\bf{u}}_{\rm{g}}^\prime T_{\rm{g}}^\prime  =  - {{\boldsymbol{\alpha }}_{{\rm{PT}}}} \cdot \nabla {T_{\rm{g}}}$, where the diagonal tensor $\boldsymbol {\alpha}_\text{PT}= \rm{diag}(\alpha_\text{pa},\alpha_\text{pe},\alpha_\text{pe})$. Like the PTRS tensor, the axis orientation of $\boldsymbol{\alpha}_{\text{PT}}$ is a local coordinate system with respect to the slip velocity, which needs to be converted to the global Cartesian coordinate system using Gram-Schmidt method. $M_\text{h}$ and $M_\text{r,m}$ in Equation \ref{e3-34} are gas-particle convection heat transfer rate and heat transfer caused by mass transfer, respectively, which can be calculated by:
\begin{equation}\label{e3-38}
{M_{\rm{h}}} + {M_{{\rm{r}},{\rm{m}}}} =  - \frac{1}{{{V_{{\rm{cell}}}}}}\sum\nolimits_{pi{\rm{ = 1}}}^{{n_{{\rm{p,cell}}}}} {({Q_{{\rm{conv,}}pi}} + {Q_{{\rm{rm,}}pi}})}.
\end{equation}
The heat generation rate due to chemical reaction is:
\begin{equation}\label{e3-39}
{M_{\rm{r}}} =  - \sum\limits_j {{H_j}{M_{{\rm{m,}}j}}}.
\end{equation}
The gas radiation between particle and gas can be ignored due to low gas emissivity \citep{norouzi2016coupled}.

The species transport equation for gas species $j$
\begin{equation}\label{e3-40}
\frac{\partial }{{\partial t}}({\varepsilon _{\rm{g}}}{\rho _{\rm{g}}}{Y_j}) + \nabla  \cdot ({\varepsilon _{\rm{g}}}{\rho _{\rm{g}}}{{\bf{u}}_{\rm{g}}}{Y_j}) = \nabla  \cdot ({\varepsilon _{\rm{g}}}{\rho _{\rm{g}}}{D_j}\nabla {Y_j}) + \nabla  \cdot ({\varepsilon _{\rm{g}}}{\rho _{\rm{g}}}{{\bf{D}}_{{\rm{PT}}}} \cdot \nabla {Y_j}) + {M_{{\rm{m,}}j}},
\end{equation}
where the first term on the right side of the equation represents the molecular diffusion, and the second term is the pseudo-turbulence diffusion term. $Y_j$ the mass fraction of species $j$, and the molecular diffusion coefficient of $j$ in gas phase can be obtained using Equation \ref{e2-27}. $\bf{{D}_\text{PT}}$ represents the pseudo-turbulence diffusion coefficient, whose value is equal to the tensor $\boldsymbol{\alpha}_\text{PT}$ \citep{peng2019implementation}.

\subsection{Coarse-grained CFD-DEM method for reacting flow}\label{s3-3}
The fluidized bed reactor used for iron ore reduction contains a large number of fine particles, and the complete conversion of iron ore to metallic iron generally takes several minutes to tens of minutes, which leads to the high time cost of simulating the complete conversion cycle of particles using traditional CFD-DEM method. In this study, we adopted a coarse-graining strategy based on EMMS-DPM (Energy Minimization Multi-Scale-Discrete Particle Method) \citep{Lu2014EMMS}. In this method, a small number of coarse-grained particles (CGPs) can be used instead of real particles to simulate large-scale systems for a long time. Coarse-grained method has been widely used in simulating gas-solid flow and heat transfer \citep{lan2020long,Lu2014EMMS,lu2016computer,lan2021simulation,lu2017extension}, but it is less used in mass transfer and chemical reaction \citep{hu2019advances,yu2021coarse,du2023coarse}. In present work, the coarse-grained method for heat transfer proposed in \cite{lu2017extension} was adopted, and the mass transfer and reaction model of CGPs were derived from it. Readers can refer to \cite{Lu2014EMMS} for the details of the coarse-grained model of particle-particle interaction, particle-wall interaction and gas-particle interphase momentum transfer. It is assumed that the density of CGP is equal to the density of real particle in present study, that is, the mass of coarse particle is $m_\text{CGP} = k^3m_\text{p}$, where $m_\text{p}$ is the mass of real particle and $k$
($k=d_\text{CGP}/d_\text{p}$) is the coarse-graining ratio. However, in the original EMMS-DPM method, there are voids inside CGP particle, and the void fraction is set as the voidage at the minimum fluidization condition. More discussions about this specific point can be found in \cite{lan2021simulation}.

The convective heat transfer rate between gas and the CGP is:
\begin{equation}\label{e3-41}
{Q_{{\rm{conv}},{\rm{CGP}}}} = {k^3}{Q_{{\rm{conv}}}}.
\end{equation}
Lu \cite{lu2017extension} deduced the relationship between the heat conduction rate of real particles with size $d_p$ and the heat conduction rate of CGPs with size $d_\text{CGP}$ as follows:
\begin{equation}\label{e3-42}
Q_{{\rm{rp}}}^{{\rm{pp}}} = \frac{{Q_{{\rm{CGP}}}^{{\rm{pp}}}}}{{{k^{9/4}}}}\sqrt {\frac{{f({e_{\rm{p}}})}}{{f({e_{{\rm{CGP}}}})}}},
\end{equation}
where $Q_{{\rm{CGP}}}^{{\rm{pp}}}$ is calculated by using Equation \ref{e3-14}, and the function of restitution coefficient is expressed as follows:
\begin{equation}\label{e3-43}
f(e) = {e^{\frac{{ - \zeta }}{{\sqrt {1 - {\zeta ^2}} \cos \zeta }}}},
\end{equation}
\begin{equation}\label{e3-44}
\zeta  = \frac{{ - \ln e}}{{\sqrt {{\pi ^2} + {{(\ln e)}^2}} }},
\end{equation}
\begin{equation}\label{e3-45}
{e_{{\rm{CGP}}}}{\rm{ = }}\sqrt {1 + k(e_{\rm{p}}^2 - 1)}.
\end{equation}
Then, the particle-particle heat conduction rate in the coarse-grained system can be written as:
\begin{equation}\label{e3-46}
Q_{{\rm{CGP}}}^{{\rm{pp'}}} = {k^3}Q_{{\rm{rp}}}^{{\rm{pp}}} = {k^{3/4}}\sqrt {\frac{{f({e_{\rm{p}}})}}{{f({e_{{\rm{CGP}}}})}}} Q_{{\rm{CGP}}}^{{\rm{pp}}}.
\end{equation}
The relationship between the heat conduction rate of real particle-fluid-particle with size $d_p$ and the heat conduction rate of coarse-grained particle-fluid-particle with size $d_\text{CGP}$ as follows \citep{lu2017extension}:
\begin{equation}\label{e3-47}
Q_{{\rm{rp}}}^{{\rm{pfp}}} = Q_{{\rm{CGP}}}^{{\rm{pfp}}}/k,
\end{equation}
where $Q_{{\rm{CGP}}}^{{\rm{pfp}}}$ is calculated by using Equation \ref{e3-15}. Therefore, the particle-fluid-particle heat conduction rate in the coarse-grained system can be written as:
\begin{equation}\label{e3-48}
Q_{{\rm{CGP}}}^{{\rm{pfp'}}} = {k^3}Q_{{\rm{rp}}}^{{\rm{pfp}}} = {k^2}Q_{{\rm{CGP}}}^{{\rm{pfp}}}.
\end{equation}
Similar to the contact heat transfer between CGPs, the heat transfer rate between the CGP and wall can be calculated by the following equations:
\begin{equation}\label{e3-48-1}
Q_{{\rm{CGP}}}^{{\rm{pw'}}} = {k^{3/4}}\sqrt {\frac{{f({e_{\rm{p}}})}}{{f({e_{{\rm{CGP}}}})}}} Q_{{\rm{CGP}}}^{{\rm{pw}}},
\end{equation}
\begin{equation}\label{e3-48-2}
Q_{{\rm{CGP}}}^{{\rm{pfw'}}} = {k^2}Q_{{\rm{CGP}}}^{{\rm{pfw}}},
\end{equation}
where $Q_{{\rm{CGP}}}^{{\rm{pw}}}$ and $Q_{{\rm{CGP}}}^{{\rm{pfw}}}$ are calculated by Equation \ref{e3-15-1} and \ref{e3-15-2}, respectively. The heat generation rate due to chemical reaction for each CGP is
\begin{equation}\label{e3-49}
{Q_{{\rm{r,CGP}}}} = {k^3}{Q_{\rm{r}}}.
\end{equation}
The heat transfer rate caused by the interphase mass transfer for each CGP is
\begin{equation}\label{e3-50}
{Q_{{\rm{rm,CGP}}}} = {k^3}{Q_{{\rm{rm}}}}.
\end{equation}
The radiation heat flow rate of single CGP from the environment is
\begin{equation}\label{e3-51}
{Q_{{\rm{rad,CGP}}}} = {k^3}{Q_{{\rm{rad}}}}.
\end{equation}
The mass transfer rate between the reduction gas $j$ and the CGP is
\begin{equation}\label{e3-52}
{\dot m_{j,{\rm{CGP}}}} = {k^3}{\dot m_j}.
\end{equation}

\subsection{Thermophysical properties of reacting gases}\label{s3-4}
Gas under normal pressure and high temperature condition can be regarded as an ideal gas, and its density is described by the ideal gas equation of state:
\begin{equation}\label{e3-53}
{\rho _{\rm{g}}} = \frac{{{M_{\rm{g}}}{p_{\rm{g}}}}}{{{R_{\rm{g}}}{T_{\rm{g}}}}}.
\end{equation}
The relationship between the viscosity of gas component $j$ and the temperature is as follows
\begin{equation}\label{e3-55}
{\mu _{{\rm{g,}}j}} = {a_1} + {a_2}{T_{\rm{g}}} + {a_3}T_{\rm{g}}^2.
\end{equation}
The coefficient $a_i$ is given in Table \ref{tab c2} (scatter data from the National Institute of Standards and Technology (https://webbook.nist.gov/chemistry/), and the coefficient is obtained by polynomial fitting). The gas phase viscosity is the mass weighting of the viscosity of each species, i.e. \citep{doble2007perry}
\begin{equation}\label{e3-56}
{\mu _{\rm{g}}} = \sum {{Y_j}{\mu _{{\rm{g,}}j}}}.
\end{equation}
The specific heat of gas species $j$ can be calculated by Equation \ref{e3-23}, and the correlation coefficients are given in Table \ref{tab c1}. The specific heat of gas phase is the mass weighting of the specific heat of each species, i.e. \citep{niksiar2010design}
\begin{equation}\label{e3-57}
{C_{p,{\rm{g}}}} = \sum {{Y_j}{C_{p,j}}}.
\end{equation}
The thermal conductivity of gas species $j$ is related to temperature as follows
\begin{equation}\label{e3-58}
{\lambda _{{\rm{g,}}j}} = {a_1} + {a_2}{T_{\rm{g}}} + {a_3}T_{\rm{g}}^2 + {a_4}T_{\rm{g}}^3 + {a_5}T_{\rm{g}}^4.
\end{equation}
The coefficient $a_i$ is given in Table \ref{tab c3}. The thermal conductivity of gas phase is the mass weighting of the thermal conductivity of each species, i.e. \citep{doble2007perry}
\begin{equation}\label{e3-59}
{\lambda _{\rm{g}}} = \sum {{Y_j}{\lambda _{{\rm{g,}}j}}}.
\end{equation}

\section{Numerical method and simulation setup}\label{s4}
\begin{figure}
\centerline{\includegraphics[width=0.8\textwidth]{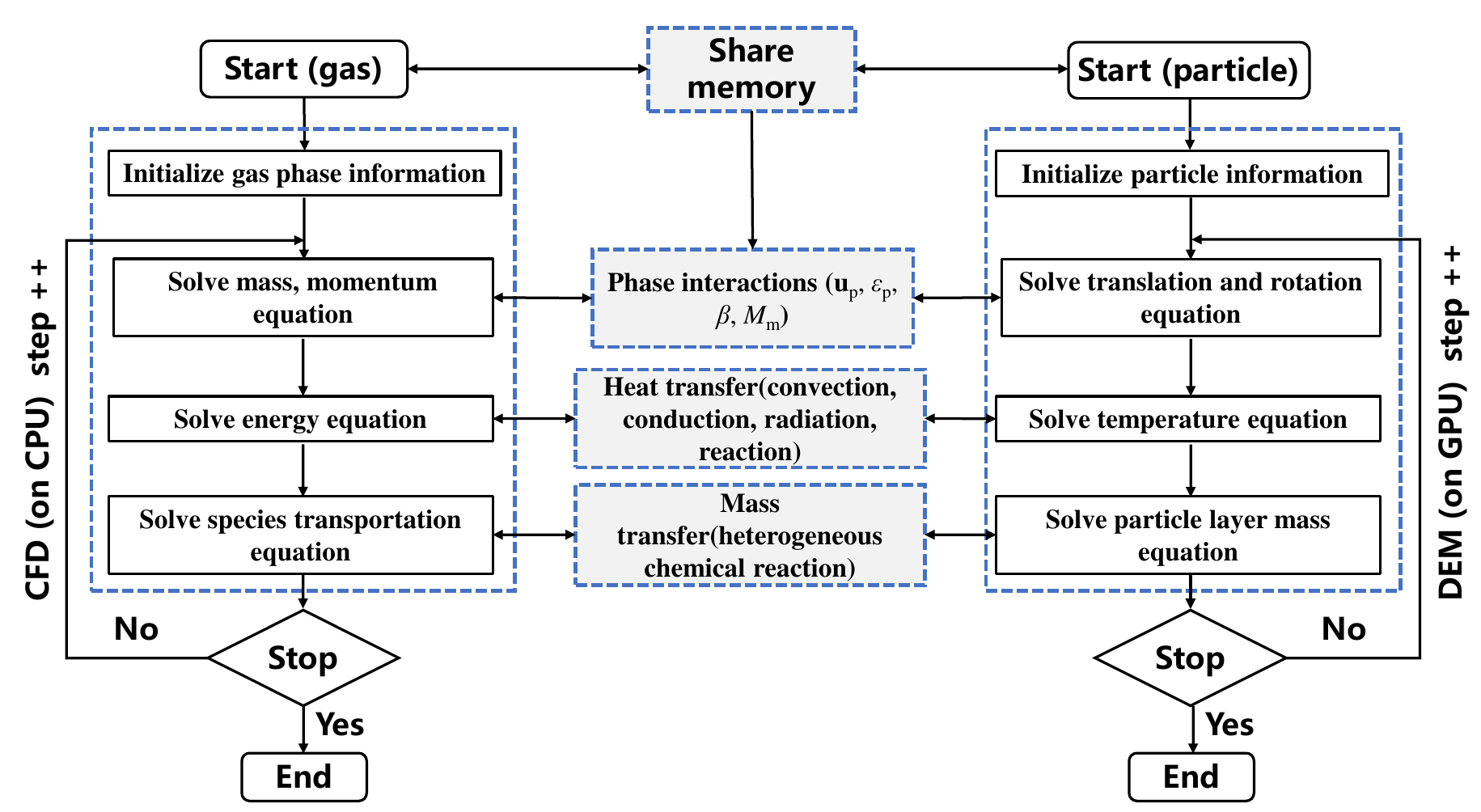}}
\caption{Flow chart of the CFD and DEM information exchange.}\label{fig5}
\end{figure}

In our previous studies, the hydrodynamic, heat and mass transfer behavior as well as catalytic reactions in gas-solid reactors have been successfully simulated using the CFD-DEM-IBM method \citep{lan2022cfd,zhao2020Acomputational,zhao2020cfd,zhao2022cartesian,zhao2022euler,lan2023critical,li2021direct,li2022cfd}. This study focuses on the implementation of the iron ore reduction reaction model on the CFD-DEM framework, so the details of IBM method are not given here again, which can be found in previous studies \cite{zhao2020Acomputational,zhao2022cartesian}. The reduction reaction is accompanied by complex flow, mass and heat transfer, in which the gas and particles exchange information frequently. This is realized via sharing memory to establish a bridge of two-phase physical data communication. As shown in Figure \ref{fig5}, gas and particle information are obtained from the central processing unit (CPU) and graphics processing unit (GPU) respectively. The interaction information of the two phases is calculated on the GPU side and transmitted to the gas phase through shared memory. Meanwhile, the gas phase information in the grid is calculated at the CPU side and passed to the particle phase through the shared memory. It should be emphasized that the exchange of information between CFD and DEM does not take place at every step, but the coupling between CFD and DEM is completed every n ($n= \Delta t_{\text{CFD}}/\Delta t_{\text{DEM}}$) step. The solvers of gas and particles complete the evolution of each physical quantity in time and space through their respective iterative operations. The particle velocity and temperature are updated at each DEM step, while the mass of each product layer of the particle is updated only when the two phases are coupled.

In the CFD-DEM simulation of chemical reactions, large interphase source terms (momentum, heat, and mass) may cause instability or even divergence in the solution of gas phase fields (velocity, temperature, and species mass fraction), present study adopts the semi-implicit treatment of source terms to avoid this situation.

\subsection{Semi-implicit treatment of source terms in momentum and energy equation}\label{s4-1}
The momentum transfer source term in Equation \ref{e3-26} can be expressed as
\begin{equation}\label{e4-1}
{{\bf{M}}_{\rm{F}}} = {K_{g{\rm{p}}}}{{\bf{u}}_{\rm{g}}} - {K_{g{\rm{p}}}}{{\bf{u}}_{\rm{p}}}.
\end{equation}
Thus, the momentum equation is rewritten as
\begin{equation}\label{e4-2}
\frac{\partial }{{\partial t}}({\varepsilon _{\rm{g}}}{\rho _{\rm{g}}}{{\bf{u}}_{\rm{g}}}) + \nabla  \cdot ({\varepsilon _{\rm{g}}}{\rho _{\rm{g}}}{{\bf{u}}_{\rm{g}}}{{\bf{u}}_{\rm{g}}}) + \nabla  \cdot \left( {{\varepsilon _{\rm{g}}}{\rho _{\rm{g}}}{\bf{u}}_{\rm{g}}^\prime {\bf{u}}_{\rm{g}}^\prime } \right) - {K_{g{\rm{p}}}}{{\bf{u}}_{\rm{g}}} =  - {\varepsilon _{\rm{g}}}\nabla p + \nabla  \cdot ({\varepsilon _{\rm{g}}}{{\bf{\tau }}_{\rm{g}}}) + {\varepsilon _{\rm{g}}}{\rho _{\rm{g}}}{\bf{g}} - {K_{g{\rm{p}}}}{{\bf{u}}_{\rm{p}}},
\end{equation}
where ${K_{g{\rm{p}}}}{{\bf{u}}_{\rm{g}}}$ and ${K_{g{\rm{p}}}}{{\bf{u}}_{\rm{p}}}$ are implicit and explicit source terms, respectively. When the temperature difference between gas and particle is high, it is necessary to transform the convection heat source term in the temperature Equation \ref{e3-34} as follows
\begin{equation*}\label{e4-3}
\begin{aligned}
{M_{\rm{h}}} &=  - \frac{1}{{{V_{{\rm{cell}}}}}}\sum\nolimits_{pi{\rm{ = 1}}}^{{n_{{\rm{p,cell}}}}} {{h_{pi}}{A_{pi}}({T_{\rm{g}}} - {T_{pi}})} \\
 &=  - \frac{1}{{{V_{{\rm{cell}}}}}}\left( {{T_{\rm{g}}}\sum\nolimits_{pi{\rm{ = 1}}}^{{n_{{\rm{p,cell}}}}} {{h_{pi}}{A_{pi}}}  - \sum\nolimits_{pi{\rm{ = 1}}}^{{n_{{\rm{p,cell}}}}} {{h_{pi}}{A_{pi}}{T_{pi}}} } \right)\\
 &=  - \frac{{\sum\nolimits_{pi{\rm{ = 1}}}^{{n_{{\rm{p,cell}}}}} {{h_{pi}}{A_{pi}}} }}{{{V_{{\rm{cell}}}}}}{T_{\rm{g}}} + \frac{{\sum\nolimits_{pi{\rm{ = 1}}}^{{n_{{\rm{p,cell}}}}} {{h_{pi}}{A_{pi}}{T_{pi}}} }}{{{V_{{\rm{cell}}}}}}\\
 &=  - {H_0}{T_{\rm{g}}} + {H_1},
\end{aligned}
\end{equation*}
where the coefficients $H_0$ and $H_1$ are respectively
\begin{equation}\label{e4-4}
{H_0} = \frac{{\sum\nolimits_{pi{\rm{ = 1}}}^{{n_{{\rm{p,cell}}}}} {{h_{pi}}{A_{pi}}} }}{{{V_{{\rm{cell}}}}}},
\end{equation}
\begin{equation}\label{e4-5}
{H_1} = \frac{{\sum\nolimits_{pi{\rm{ = 1}}}^{{n_{{\rm{p,cell}}}}} {{h_{pi}}{A_{pi}}{T_{pi}}} }}{{{V_{{\rm{cell}}}}}}.
\end{equation}
In this way, the energy equation can be rewritten as
\begin{equation}\label{e4-6}
\frac{\partial }{{\partial t}}({\varepsilon _{\rm{g}}}{\rho _{\rm{g}}}{C_{p,{\rm{g}}}}{T_{\rm{g}}}) + \nabla  \cdot ({\varepsilon _{\rm{g}}}{\rho _{\rm{g}}}{{\bf{u}}_{\rm{g}}}{C_{p,{\rm{g}}}}{T_{\rm{g}}}) + \nabla  \cdot ({\varepsilon _{\rm{g}}}{\rho _{\rm{g}}}{C_{p,{\rm{g}}}}{\bf{u}}_{\rm{g}}^\prime T_{\rm{g}}^\prime ) + {H_0}{T_{\rm{g}}} = \nabla  \cdot ({\varepsilon _{\rm{g}}}{\lambda _{\rm{g}}}\nabla {T_{\rm{g}}}) + {H_1} + {M_{\rm{r}}} + {M_{{\rm{r,m}}}}.
\end{equation}
It should be noted that in the CFD simulation process, we found that the values of the reaction heat source term and the heat transfer source term caused by mass transfer are similar, and the sign is opposite, so the two terms remain explicit.

\subsection{Semi-implicit treatment of mass source term in species transport equation}\label{s4-2}
For reducing gas $j$, mass transfer rate on a single particle scale
\begin{equation}\label{e4-7}
\begin{split}
{{\dot m}_{j,pi}} &= \frac{{{\eta _j}{A_{pi}}}}{{{T_{pi}}}}\frac{{{M_j}{p_{\rm{g}}}}}{{{R_{\rm{g}}}}}{\left[ {{v_1}({x_j} - x_{{\rm{eq}},j}^{{\rm{hm}}}) + {v_2}({x_j} - x_{{\rm{eq}},j}^{{\rm{mw}}}) + {v_3}({x_j} - x_{{\rm{eq}},j}^{{\rm{wFe}}})} \right]_j}\\
 &= \frac{{{\eta _j}{A_{pi}}}}{{{T_{pi}}}}\frac{{{M_j}{p_{\rm{g}}}}}{{{R_{\rm{g}}}}}{\left[ {({v_1} + {v_2} + {v_3}){x_j} - ({v_1}x_{{\rm{eq}},j}^{{\rm{hm}}} + {v_2}x_{{\rm{eq}},j}^{{\rm{mw}}} + {v_3}x_{{\rm{eq}},j}^{{\rm{wFe}}})} \right]_j},
\end{split}
\end{equation}
where ${v_1} = v_1^{{\rm{hm}}} + v_1^{{\rm{mw}}} + v_1^{{\rm{wFe}}}$, ${v_2} = v_2^{{\rm{hm}}} + v_2^{{\rm{mw}}} + v_2^{{\rm{wFe}}}$, ${v_3} = v_3^{{\rm{hm}}} + v_3^{{\rm{mw}}} + v_3^{{\rm{wFe}}}$. Let ${E_j} = {\eta _j}{A_{pi}}/{T_{pi}}$, ${Q_j} = {v_1} + {v_2} + {v_3}$, ${S_j} = {v_1}x_{{\rm{eq}},j}^{{\rm{hm}}} + {v_2}x_{{\rm{eq}},j}^{{\rm{mw}}} + {v_3}x_{{\rm{eq}},j}^{{\rm{wFe}}}$, then, we have
\begin{equation}\label{e4-8}
\begin{split}
{{\dot m}_{j,pi}} &= {E_j}{M_j}\left[ {{Q_j}{x_j} - {S_j}} \right]\frac{{{p_{\rm{g}}}}}{{{R_{\rm{g}}}}}\\
 &= {E_j}{M_j}{Q_j}{x_j}\frac{{{p_{\rm{g}}}}}{{{R_{\rm{g}}}}} - {E_j}{S_j}\frac{{{M_j}}}{{{R_{\rm{g}}}}}{p_{\rm{g}}}\\
 &= {E_j}{M_j}{Q_j}\left( {\frac{{{{x'}_j}}}{{{X_j}}}} \right)\frac{{{p_{\rm{g}}}}}{{{R_{\rm{g}}}}} - {E_j}{S_j}\frac{{{M_j}}}{{{R_{\rm{g}}}}}{p_{\rm{g}}}\\
 &= {E_j}{M_j}{Q_j}\left( {\frac{{{M_{\rm{g}}}{Y_j}/{M_j}}}{{{X_j}}}} \right)\frac{{{p_{\rm{g}}}}}{{{R_{\rm{g}}}}} - {E_j}{S_j}\frac{{{M_j}}}{{{R_{\rm{g}}}}}{p_{\rm{g}}}\\
 &= \frac{{{E_j}{Q_j}{\rho _{\rm{g}}}{T_{\rm{g}}}}}{{{X_j}}}{Y_j} - {E_j}{S_j}\frac{{{M_j}}}{{{R_{\rm{g}}}}}{p_{\rm{g}}},
\end{split}
\end{equation}
where ${X_j} = {x'_{{{\rm{H}}_2}}} + {x'_{{{\rm{H}}_2}{\rm{O}}}}$. By substituting Equations \ref{e4-8} and \ref{e3-25} into Equation \ref{e3-40}, the species transport equation is rewritten as
\begin{equation}\label{e4-9}
\frac{\partial }{{\partial t}}({\varepsilon _{\rm{g}}}{\rho _{\rm{g}}}{Y_j}) + \nabla  \cdot ({\varepsilon _{\rm{g}}}{\rho _{\rm{g}}}{{\bf{u}}_{\rm{g}}}{Y_j}) - {C_0}\frac{{{\rho _{\rm{g}}}{T_{\rm{g}}}}}{{{X_j}}}{Y_j} = \nabla  \cdot ({\varepsilon _{\rm{g}}}{\rho _{\rm{g}}}{D_j}\nabla {Y_j}) + \nabla  \cdot ({\varepsilon _{\rm{g}}}{\rho _{\rm{g}}}{{\bf{D}}_{{\rm{PT}}}} \cdot \nabla {Y_j}) + {C_1}{p_{\rm{g}}}.
\end{equation}
Where $C_0$ and $C_1$ are respectively
\begin{equation}\label{e4-10}
{C_0} = \frac{{\sum\nolimits_{pi{\rm{ = 1}}}^{{n_{{\rm{p,cell}}}}} {{E_j}{Q_j}} }}{{{V_{{\rm{cell}}}}}},
\end{equation}
\begin{equation}\label{e4-11}
{C_1} =  - \frac{{{M_j}}}{{{R_{\rm{g}}}}}\frac{{\sum\nolimits_{pi{\rm{ = 1}}}^{{n_{{\rm{p,cell}}}}} {{E_j}{S_j}} }}{{{V_{{\rm{cell}}}}}}.
\end{equation}

\subsection{Simulation parameters and settings}\label{s4-2}
\begin{figure}
\centerline{\includegraphics[width=0.5\textwidth]{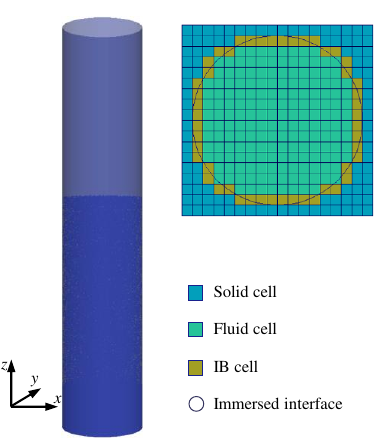}}
\caption{Schematic illustration of reduction reactor and grid distribution (top view).}\label{fig6}
\end{figure}

In order to verify the correctness of the CFD-DEM-IBM method coupled with the USCM model, the reduction processes of magnetite and hematite were respectively simulated and compared with the experimental results of \cite{du2022relationship} and \cite{spreitzer2019iron}. Magnetite is reduced in a bubbling fluidized bed with an inner diameter of 20 mm and a height of 60 mm, and the hematite reduction reactor has an inner diameter of 68 mm and a height of 200 mm. Figure \ref{fig6} shows the schematic illustration of the reactor and the fluid grid distribution. Table \ref{tab3} lists the simulation parameters and physical properties of the particle and gases. It should be pointed out that the magnetite and hematite particles used in the experiment both are irregular, so Equation \ref{e3-5-2} are used to calculate the drag coefficient. In addition, the discrete scheme of each term in the gas conservation equation are shown in Table \ref{tab4}.

\begin{table}[]
\centering
\caption{Parameters and physical properties used in present simulations.}\label{tab3}
\begin{tabular}{cccc}
\hline
                          & Parameter                         & Bed I                  & Bed II   \\ \hline
\multirow{18}{*}{Particle}& Iron ore type           & magnetite                          &    hematite         \\
                          & Initial efficient density, $\rho_\text{p0}$ (kg/m$^3$)           & 3720                          & 3500            \\
                          & True density, $\rho_\text{ture}$ (kg/m$^3$)                & \multicolumn{2}{c}{5300($\rm{Fe_2O_3}$), 5180($\rm{Fe_3O_4}$), 5390(FeO), 7900(Fe)}                        \\
                          & Mean diameter of real particle, $d_\text{p}$ (mm)                & 0.1              & 0.35            \\
                          & Number of real particle               & 5140800              & 5095837            \\
                          & Coarse-graining ratio, $k$               & 2.8              & 3.0            \\
                          & Young's modulus, $Y_\text{p}$ (Pa)       &\multicolumn{2}{c}{$1\times10^{{\rm{7}}}$}       \\
                          & Restitution coefficient of real particle, $e_\text{p}$        &\multicolumn{2}{c}{0.9}            \\
                          & Sliding friction coefficient, $\mu_\text{s}$           &\multicolumn{2}{c}{0.3}            \\
                          & Rolling friction coefficient, $\mu_\text{r}$ (mm) &\multicolumn{2}{c}{0.01}            \\
                          & Poisson's ratio, $\nu_p$                   &\multicolumn{2}{c}{0.3}             \\
                          & Characteristic velocity, $u_\text{c}$ (m/s)     &\multicolumn{2}{c}{1.0}             \\
                          & Initial temperature, $T_\text{p0}$ (K)           & 943, 923, 903                           & 1073, 973   \\
                          & Initial mass fraction (wt$\%$)    & 69.94($\rm{Fe_3O_4}$), 30.06(FeO)     & 96.28($\rm{Fe_2O_3}$), 3.72(FeO)   \\
                          & Initial porosity, $\xi_0$           & 0.29                 & 0.34   \\
                          & Pore diameter (cm), $d_a$           & $9.5\times10^{{\rm{-6}}}$          & $3.3\times10^{{\rm{-6}}}$   \\
                          & Emissivity, $\eta _{\rm{p}}$           &\multicolumn{2}{c}{0.85}   \\
                          & Sphericity, $\psi$           &0.85  & 0.86   \\
                          & Time step,  $\Delta t_\text{DEM}$ (s) & $5\times10^{{\rm{-}}6}$ &$2\times10^{{\rm{-}}5}$         \\ \hline
\multirow{5}{*}{Gas}      & Inlet composition, $x_{j0}$ (vol$\%$)     & 70($\rm{H_2}$), 30($\rm{N_2}$)       & 65($\rm{H_2}$), 35($\rm{N_2}$)       \\
                          & Inlet velocity, $u_\text{g0}$ (m/s)     & 0.183, 0.179, 0.175       & 0.43, 0.39         \\
                          & Inlet temperature, $T_\text{g0}$ (K)     & 943, 923, 903                & 1073, 973    \\
                          & Operating pressure, $p_\text{go}$ (Pa)           & 101325                       & 110000      \\
                          & Grid size, $L_x\times L_y \times L_z$ (mm)          & $1.25\times 1.25 \times 1.25$                       & $4.25\times 4.25 \times 4.25$      \\
                          & Time step, $\Delta t_\text{CFD}$ (s)   &$1\times10^{{\rm{-}}4}$   & $2\times10^{{\rm{-}}4}$            \\ \hline
\multirow{3}{*}{Boundary condition}      & Wall velocity        &\multicolumn{2}{c}{No slip}  \\
                          & Wall temperature (K)        &943, 923, 903  & 1073, 973  \\
                          & Outlet Pressure (Pa)   &101325   &110000   \\  \hline
\end{tabular}
\end{table}

\begin{table}[]
\centering
\caption{Numerical scheme used in OpenFOAM.}\label{tab4}
\begin{tabular}{cc}
\hline
Term                         & Discrete scheme   \\ \hline
Time derivative     &    Euler \\
Temperature gradient        & Gauss linear            \\
Velocity gradient                        &CellLimited Gauss linear 1       \\
Pressure gradient                        & Gauss linear      \\
Convection (velocity)                        &  GausslimitedLinearV 1     \\
Convection (temperature)                        &  Gauss upwind     \\
Convection (species mass fraction)                        &  limitedLinear01 1     \\
Laplacian                        & Gauss linear corrected      \\      \hline
\end{tabular}
\end{table}

\section{Results and discussion}\label{s5}

\subsection{Reduction of magnetite}\label{s5-1}
\subsubsection{Determination of reaction kinetic parameters and their influence}\label{s5-1-1}
Reaction kinetic parameters (pre-exponential factor and activation energy) are crucial to the reduction reaction rate and iron ore conversion rate, which are generally determined by linear regression method in experiments \citep{heidari2021review,spreitzer2019iron,spreitzer2020fluidization}. Du \cite{du2022relationship} obtained the apparent activation energy of magnetite reduction in $\rm{H_2}$ atmosphere by linear regression in the experiment, but the activation energy of each specific reaction was unknown, we adopted the values given in \cite{takenaka1986mathematical} for the activation energy of reaction R2 and R3 in CFD simulations: ${E_\text{a}^{\text{mw}}}$=75345 J/mol, ${E_\text{a}^{\text{wFe}}}$=117204 J/mol. Furthermore, the method provided by \cite{takenaka1986mathematical} was used to estimate the pre-exponential factor: values considered appropriate were substituted initially, and the pre-exponential factor was constantly adjusted on the premise of keeping other operating conditions unchanged, while the change in the reduction degree of the particles was monitored. When the reduction degree is in good agreement with the experimental results, the pre-exponential factor of each sub-reaction can be determined.

\begin{figure}
\centerline{\includegraphics[width=0.7\textwidth]{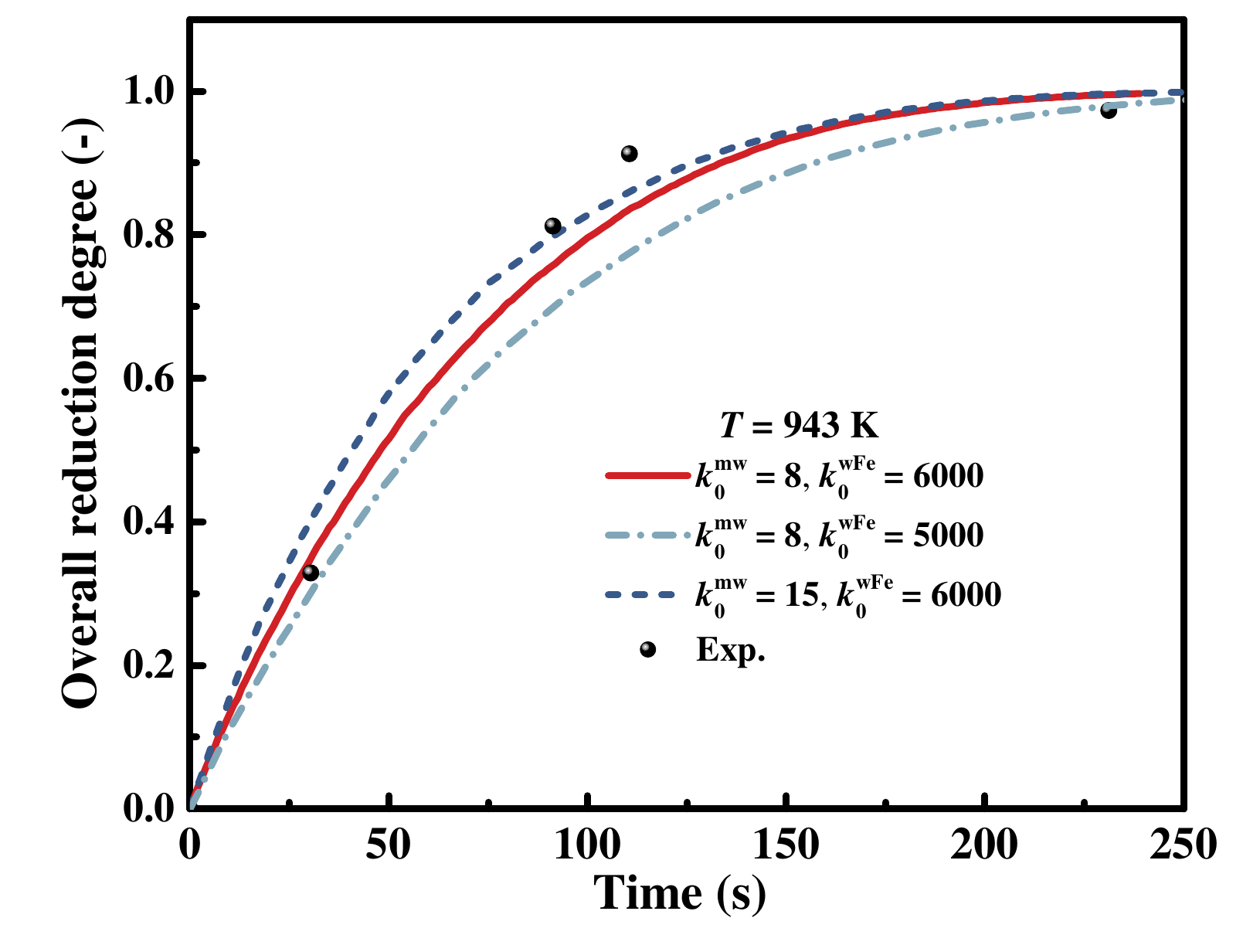}}
\caption{Effect of pre-exponential factors on magnetite reduction.}\label{fig7}
\end{figure}

As shown in Figure \ref{fig7}, when the initial values of the pre-exponential factors of the reaction from $\rm{Fe_3O_4}$ to FeO and FeO to Fe are 8 and 5000, respectively, the predicted overall reduction degree of magnetite deviates greatly from the experimental values, especially in the middle stages of the reaction. Therefore, by adjusting these two values, a set of pre-exponential factors that best agree with the experimental values are obtained successfully: ${k_0^{\text{mw}}}$=8 m/s, ${k_0^{\text{wFe}}}$=6000 m/s. The slope of the reduction curve represents the reaction rate. Before about 20 s, the particles undergo a constant rate transformation period. With the reduction of magnetite, the slope becomes smaller and the reaction rate slows down, indicating that the reaction resistance in the early stage is less than that in the later stage. Additionally, as the pre-exponential factor increases, the reaction rate constant increases, meaning that less time is required for the complete conversion of magnetite to metallic iron.

\begin{figure}
\centerline{\includegraphics[width=1.0\textwidth]{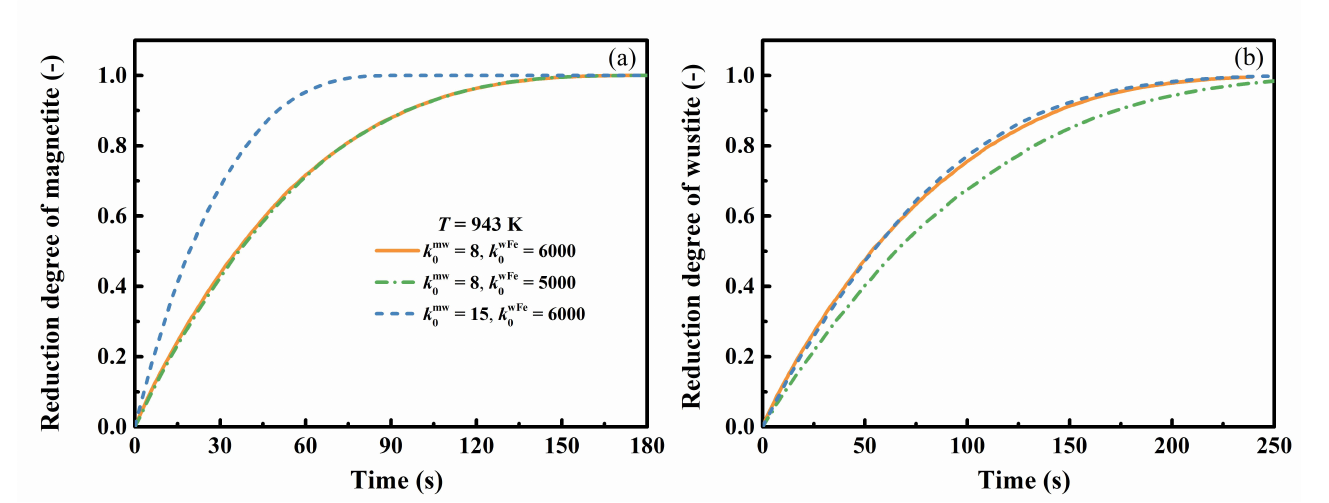}}
\caption{Effect of pre-exponential factors on fractional reduction.}\label{fig8}
\end{figure}

Figure \ref{fig8} illustrates the effect of pre-exponential factors on the reduction degree of $\rm{Fe_3O_4}$ (a) and FeO (b). It is obvious that the pre-exponential factor of the reaction $\rm{Fe_3O_4}$$\rightarrow$ FeO (${k_0^{\text{mw}}}$) has a great influence on the reduction of $\rm{Fe_3O_4}$$\rightarrow$FeO, while the pre-exponential factor of the reaction FeO$\rightarrow$Fe (${k_0^{\text{wFe}}}$) has almost no effect on the conversion of $\rm{Fe_3O_4}$. It can be seen from Figure \ref{fig8} (b) that at the initial stage of the reaction, the reduction degree of FeO depends only on ${k_0^{\text{wFe}}}$, and has almost no relationship with ${k_0^{\text{mw}}}$. With the progress of the reaction, the increase of ${k_0^{\text{mw}}}$ will slightly accelerate the conversion of FeO. The possible reason is that when ${k_0^{\text{mw}}}$ changes from 8 to 15 (m/s), the time for the complete conversion of $\rm{Fe_3O_4}$ to FeO is greatly shortened, and more hydrogen will react with FeO in the middle and late stage of the reaction, thus accelerating the reduction of FeO.

\begin{figure}
\centerline{\includegraphics[width=1.0\textwidth]{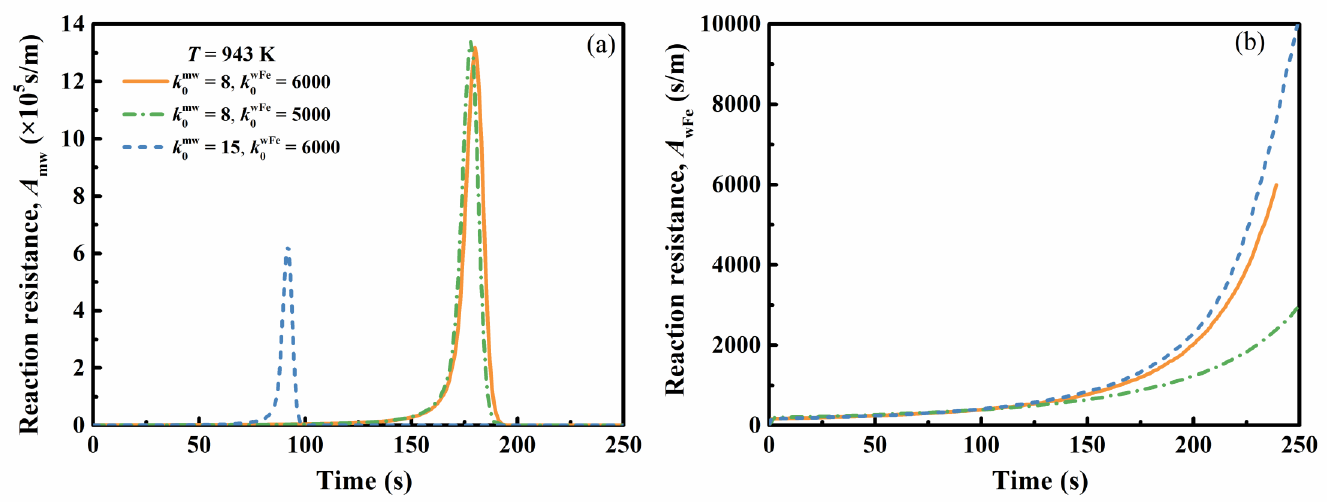}}
\caption{Effect of pre-exponential factors on the interfacial reaction resistance.}\label{fig9}
\end{figure}

In Figure \ref{fig9}, the influence of pre-exponential factors on the interfacial reaction resistance is illustrated. As the reaction progresses, the chemical reaction resistance increases. As can be seen from Figure \ref{fig9} (a), when $\rm{Fe_3O_4}$ is almost completely converted into FeO, the interfacial reaction resistance of reaction from $\rm{Fe_3O_4}$ to FeO ($A_\text{mw}$) reaches its peak value and then rapidly drops to zero, a similar phenomenon that was also observed by Kinaci\cite{kinaci2020cfd}. At this time, $\rm{Fe_3O_4}$ in the particle has been completely consumed, and the FeO formation reaction no longer occurs. According to Equation \ref{e2-12}, when the reaction temperature changes within a narrow range, the reaction rate constant and reduction degree are the most critical variables affecting the interface reaction resistance. For a given reaction rate constant, the reaction resistance increases with the increase of the reduction degree, and when the reduction degree of the particles is the same (for example, the reduction degree of $\rm{Fe_3O_4}$ corresponding to the peak value in Figure \ref{fig8} (a) is about 1), the reaction rate constant is negatively correlated with the reaction resistance. Consistent with the change in reduction degree, $A_\text{mw}$ is only related to $A_\text{mw}$, while $A_\text{wFe}$ is not only closely related to ${k_0^{\text{wFe}}}$, but is also influenced by ${k_0^{\text{mw}}}$, for possible reasons already given in the discussion about Figure \ref{fig8}.

\begin{figure}
\centerline{\includegraphics[width=1.0\textwidth]{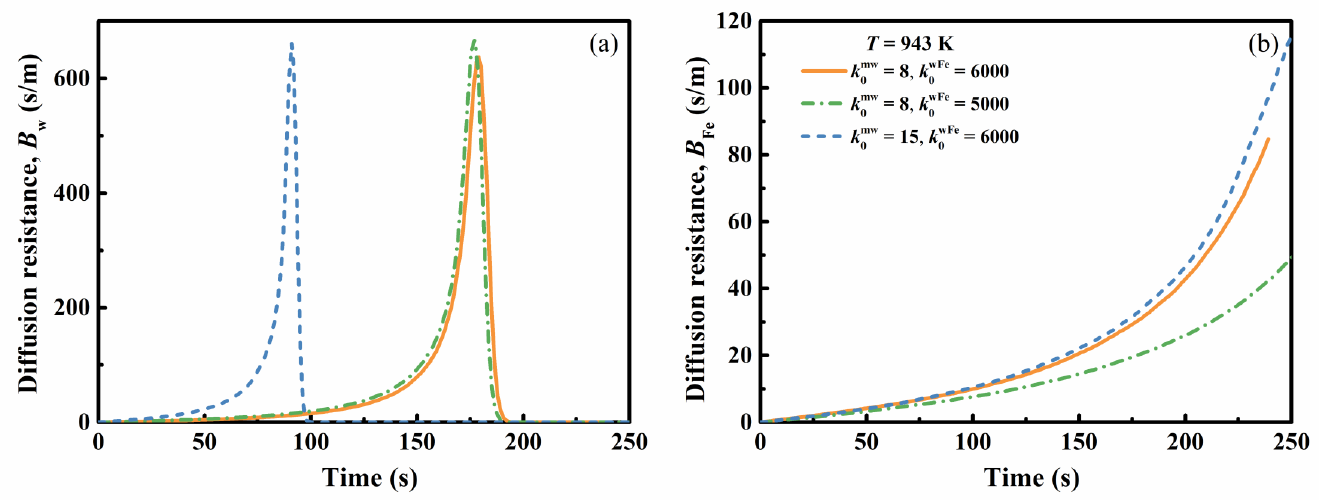}}
\caption{Effect of pre-exponential factors on the product layers diffusion resistance.}\label{fig10}
\end{figure}

The influence of pre-exponential factors on the diffusion resistance of gas through the product layer is shown in Figure \ref{fig10}. With the reduction of magnetite, the thickness of the product layer increases, and its resistance to gas diffusion increases gradually. When $\rm{Fe_3O_4}$ is completely reduced to FeO, the resistance ($B_\text{w}$) of the gas through the wustite reaches its peak. After this, $B_\text{w}$ becomes 0 because there is only one product layer of metallic iron in the particle. The resistance of the gas through the iron layer ($B_\text{Fe}$) increases until the FeO is completely consumed. The variation trend of internal diffusion resistance with the pre-exponential factor is similar to that of the reaction resistance in Figure \ref{fig10}, and the analysis will not be carried out here.

\begin{figure}
\centerline{\includegraphics[width=0.7\textwidth]{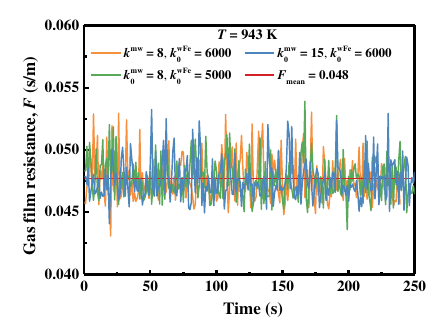}}
\caption{Effect of pre-exponential factors on gas film diffusion resistance.}\label{fig11}
\end{figure}

As shown in Figure \ref{fig11}, in the magnetite reduction process, the mass transfer resistance through gas film fluctuates around an mean value ($F_\text{mean}$=0.048 s/m), and the pre-exponential factor has little influence on it. From Equations \ref{e2-20} and \ref{e2-21}, it can be seen that the external diffusion resistance is only related to the diffusion coefficient $D_\text{eff,f}$ and the dimensionless Sherwood number $Sh$ of the reducing gas in the gas film, while the pre-exponential factor has little impact on these two gas-related variables. Therefore, the time-averaged value of the gas film resistance predicted by the three different pre-exponential factors is basically equal.

In summary, the interfacial reaction resistance is largest, and the order of each resistance is: $A_\text{mw}>A_\text{wFe}>B_\text{w}>B_\text{Fe}>F$. In other words, the chemical reaction at the interfaces is the dominant control mechanism during magnetite reduction. Most researchers \citep{nicolle1979mechanism,patisson2020hydrogen,heidari2021review} believe that a dense iron shell forms on the outer surface of the particle when wustite is reduced to metallic iron, and the diffusion of hydrogen through this shell becomes difficult, which is consistent with our findings.

\begin{figure}
\centerline{\includegraphics[width=0.8\textwidth]{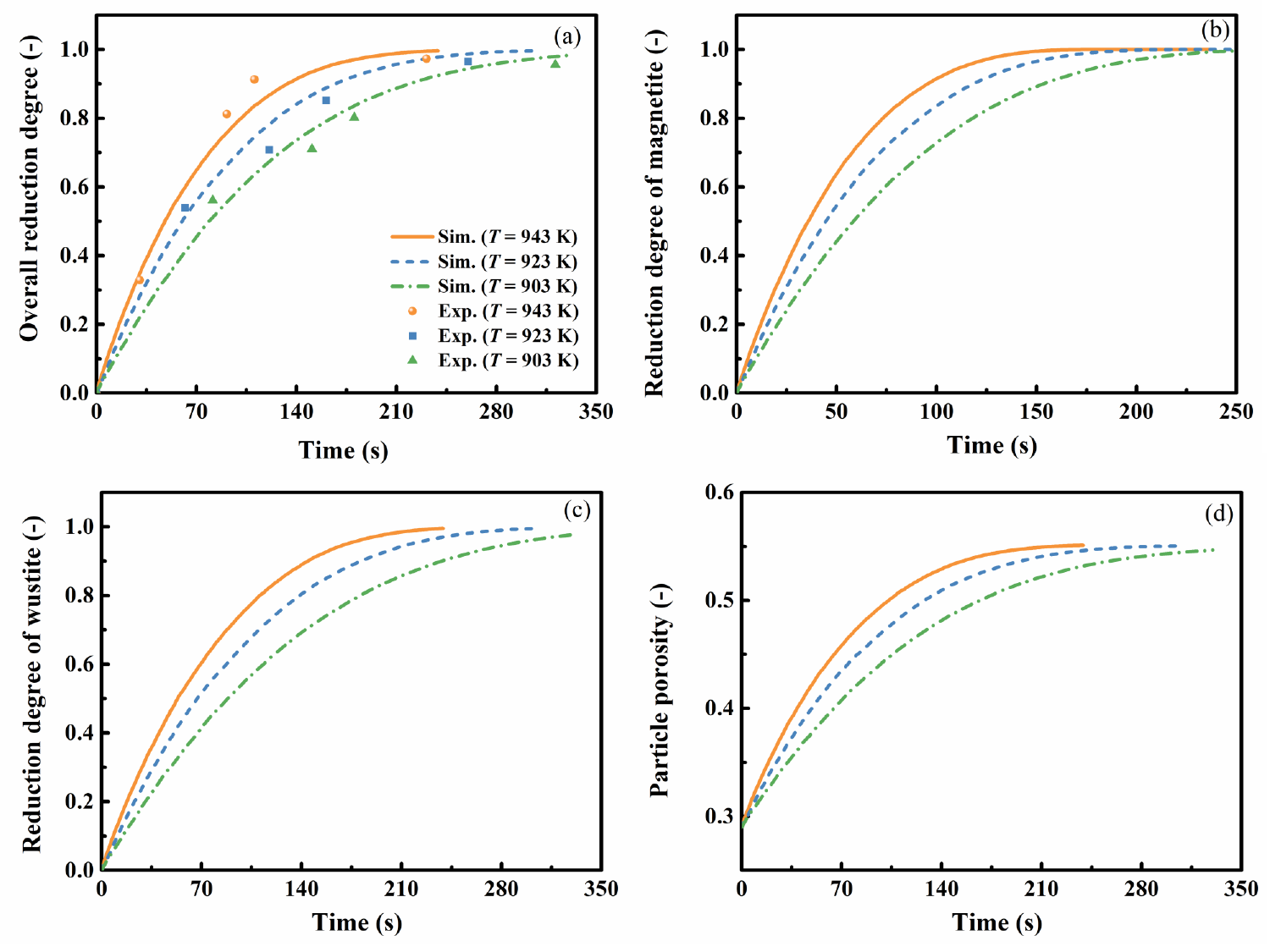}}
\caption{Effect of operating temperature on magnetite reduction.}\label{fig12}
\end{figure}

\subsubsection{Effect of temperature on magnetite reduction}\label{s5-1-2}

Figure \ref{fig12} shows the comparison of the reduction progress at different operating temperatures. The agreement between the measured and calculated values is very good at 943, 923, and 903 K, while the overall reduction degree at the final stage is less satisfactorily simulated (all overestimated). The possible reason is that our model does not take into account that the adhesion of iron to the surface of the wustite causes the reduction rate to be greatly reduced. Obviously, due to the heat absorption during magnetite reduction, the increase of operating temperature will accelerate the reaction rate, and the reduction degree and porosity of the particles will increase accordingly, but the final porosity does not change, about 0.55.

\subsection{Reduction of hematite}\label{s5-2}
\subsubsection{Determination of reaction kinetic parameters and model validation}\label{s5-2-1}

\begin{figure}
\centerline{\includegraphics[width=0.7\textwidth]{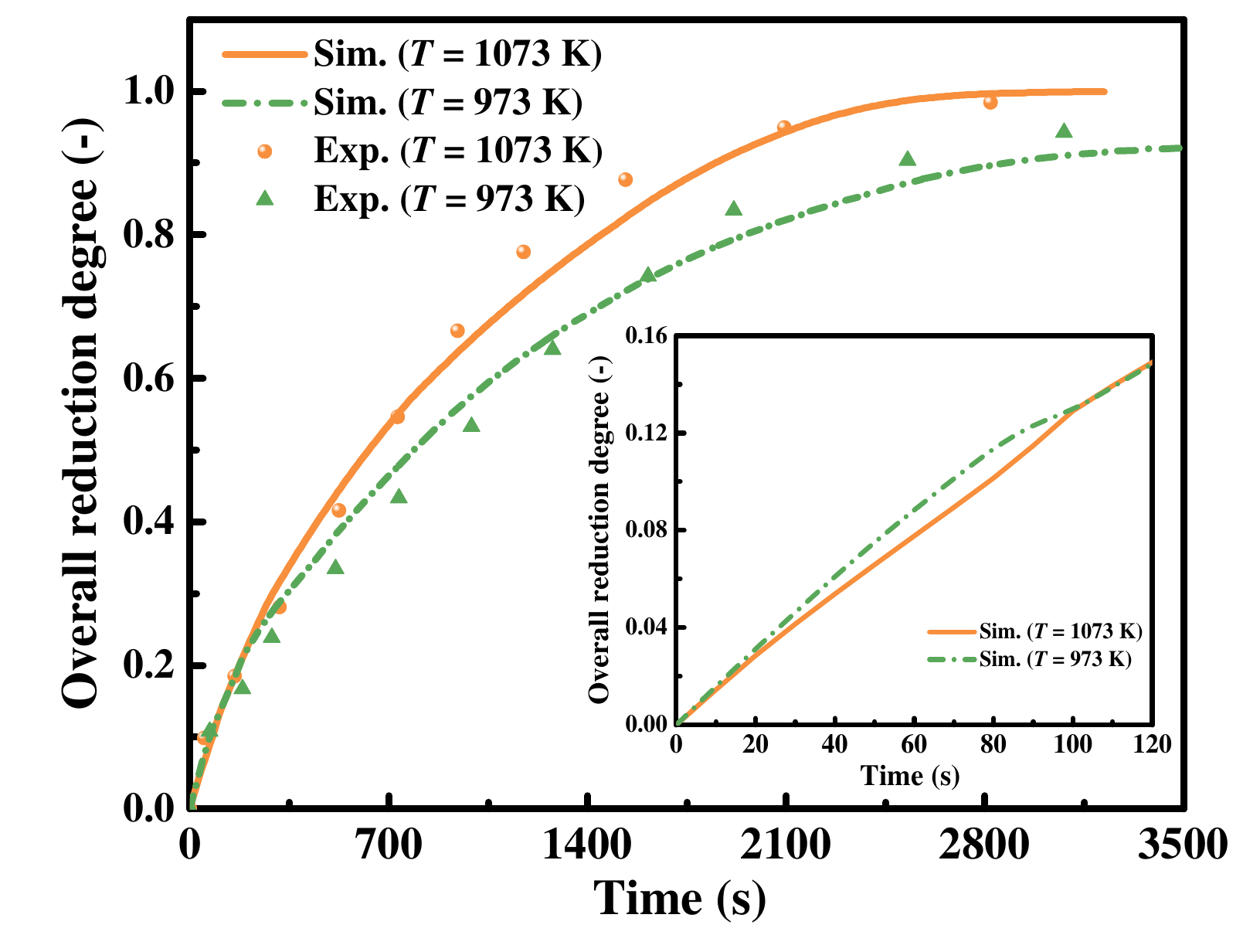}}
\caption{Effect of pre-exponential factors on hematite reduction.}\label{fig13}
\end{figure}

Spreitzer \cite{spreitzer2020development} obtained the activation energies of reactions $\rm{Fe_2O_3}$$\rightarrow$$\rm{Fe_3O_4}$, $\rm{Fe_3O_4}$$\rightarrow$FeO and FeO$\rightarrow$Fe by fitting experimental data, which were 42.4, 19.3 and 33.88 (kJ/mol), respectively. When the operating temperature is 1073 K, the method in Section \ref{s5-1-1} is still used to estimate the pre-exponential factors of each sub-reaction, and the final pre-exponential factors are 10, 0.0007, 0.022 (m/s), respectively. Figure \ref{fig13} shows the change of the overall reduction degree of hematite particles with time at 973 K, 1023 K and 1073 K. It can be seen that in a certain temperature range (973 K$\sim$1073 K), our model can accurately predict the direct reduction process of hematite in the mixed atmosphere of hydrogen and nitrogen. Hematite takes about an hour to be almost completely converted into metallic iron, while magnetite (Figure \ref{fig5}) takes less than five minutes. The reasons for this difference may be: (i) the average particle size of hematite is larger than that of magnetite, and (ii) the reaction rate constant of hematite is smaller than that of magnetite. The second reason may be dominant, and the effect of particle size on the reaction rate is actually small \citep{du2022relationship}. In the reaction from $\rm{Fe_2O_3}$ to $\rm{Fe_3O_4}$, the equilibrium concentration of hydrogen is close to zero. In the initial stage ($t<$100 s), most of the hydrogen is used to reduce $\rm{Fe_2O_3}$, and the generation of $\rm{Fe_3O_4}$ is an exothermic process, so the reduction degree increases with the drop of temperature during this period (local zoom in Figure \ref{fig13}). After $\rm{Fe_2O_3}$ is consumed, endothermic reaction occurs, and the increase in temperature is conducive to the reduction of particles. Similarly, the reduction rate is slower due to the deposition of iron on the particle surface.

\begin{figure}
\centerline{\includegraphics[width=0.8\textwidth]{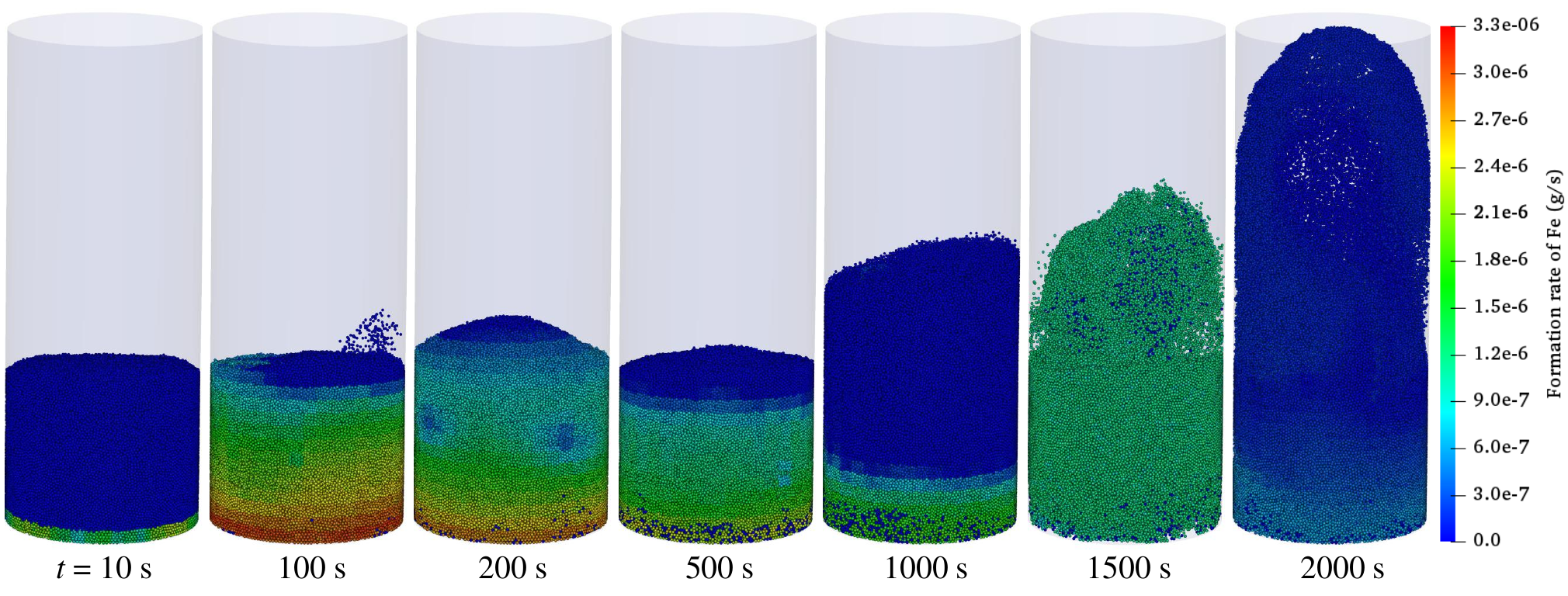}}
\caption{Evolution of iron formation rate at 1073 K.}\label{fig14}
\end{figure}
In the reduction process of hematite, metallic iron is constantly generated, and the change of the rate of iron generation in each particle over time is shown in Figure \ref{fig14}. It can be seen that the rate of iron formation in the lower part of the bed is higher than that in the upper part. At the beginning of the reaction, only the conversion of $\rm{Fe_2O_3}$ to $\rm{Fe_3O_4}$ occurs in most of the particles in the bed, and no iron is formed. After 100 s, more and more particles complete the transition from hematite to metallic iron. Until 2000 s, a small number of particles were not completely reduced. Obviously, with the progress of the reaction, the expansion height of the bed increases, and the flow regime undergoes a transition from a fixed bed to a bubbling bed. This is because the particles become lighter when they are reduced, and the resultant force of the drag and gravity on the particles increases when the operating gas velocity is unchanged. Therefore, when hematite is completely reduced to Fe, the particles will be more easily carried by bubbles to the surface of the bed.

\subsubsection{Effect of temperature on hematite reduction}\label{s5-1-2}
\begin{figure}
\centerline{\includegraphics[width=0.8\textwidth]{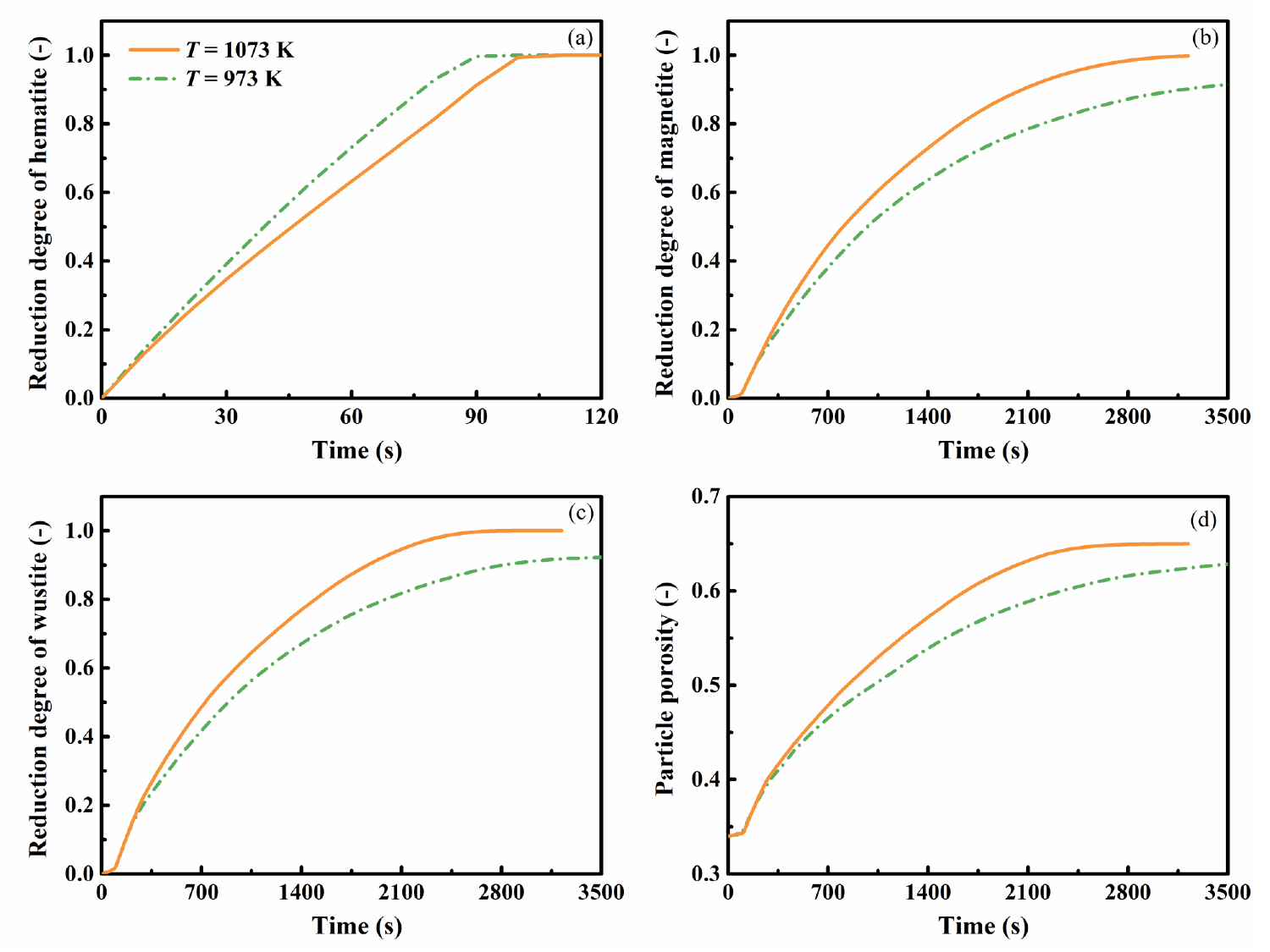}}
\caption{Effect of temperature on reduction degree and porosity.}\label{fig16}
\end{figure}

Figure \ref{fig16} gives the quantitative comparison among the changes of fractional reduction degree and particle porosity during hematite reduction at different operating temperatures. Obviously, because the reduction of $\rm{Fe_2O_3}$ to $\rm{Fe_3O_4}$ is an endothermic reaction, the reaction rate of $\rm{Fe_2O_3}$ at higher temperatures is slower, which means a low fractional reduction. Before 100 s, the R1 reaction mainly occurs in the particle, so the reduction degree of $\rm{Fe_3O_4}$ and FeO is very low. Furthermore, less reduction in particle mass at this stage leads to little change in porosity. After $\rm{Fe_2O_3}$ is completely consumed, the reduction degree of $\rm{Fe_3O_4}$ and FeO increases with the rise of temperature. When the operating temperature is 1073 K, the complete conversion of hematite to metallic iron takes about 3200 s, while at 973 K, the entire conversion process takes longer. The increase in the operating temperature accelerates the loss of oxygen and causes the porosity of the particles to increase until the end of the reaction, at which point the porosity is about 0.65. Overall, the high temperature is conducive to the reduction of iron ore with hydrogen.

\begin{figure}
\centerline{\includegraphics[width=0.8\textwidth]{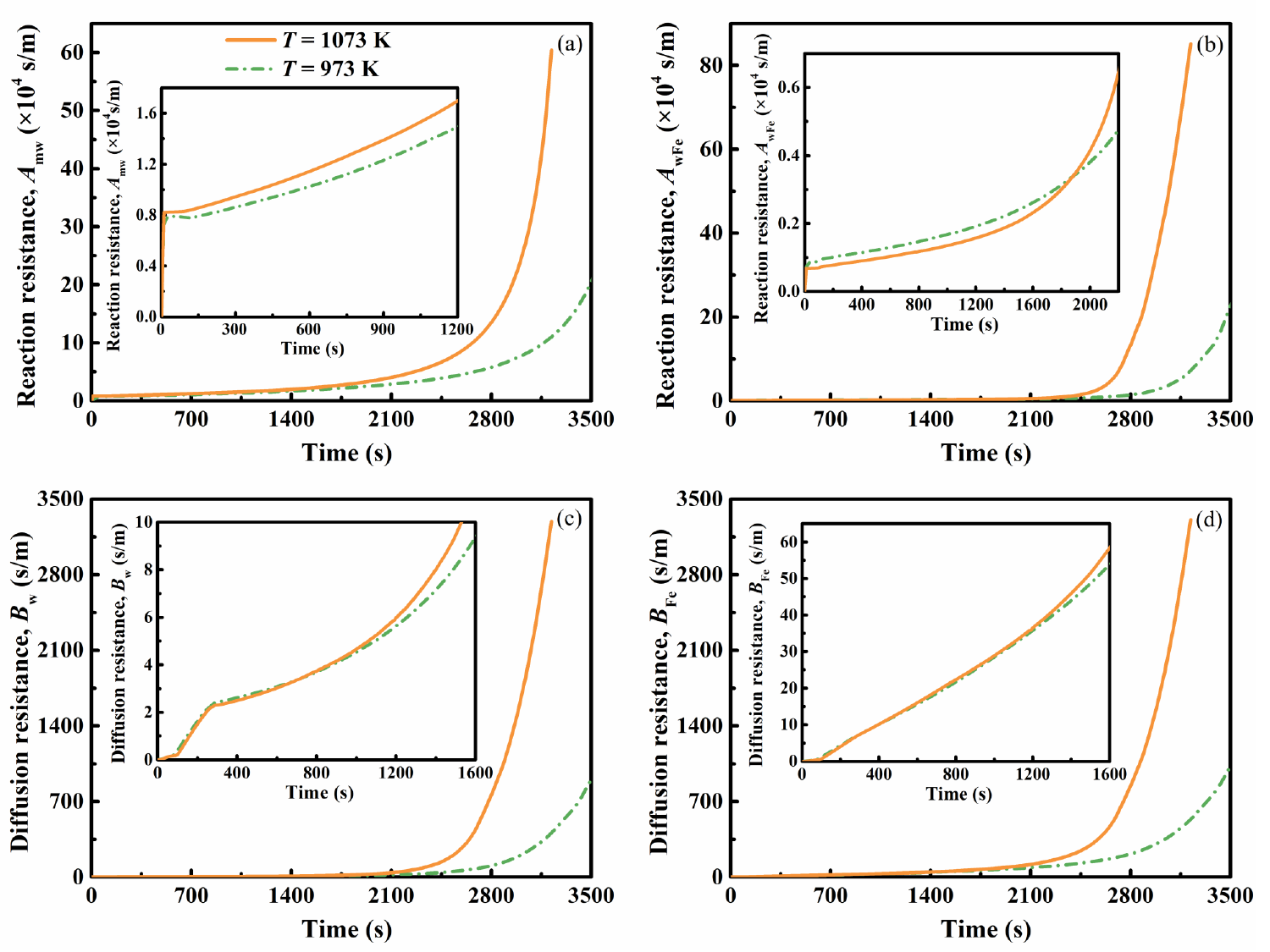}}
\caption{Effect of temperature on reaction resistance and diffusion resistance.}\label{fig17}
\end{figure}

Figure \ref{fig17} shows the influence of temperature on interfacial chemical reaction resistance and gas diffusion resistance through the product layers during hematite reduction. With the progress of the reaction, both kinds of resistance increased, and the resistance increased sharply at the end of the reaction, which was consistent with the trend of resistance change of magnetite particles in Section \ref{s5-1}. The influence trend of temperature on each resistance in the early stage of reaction is inconsistent (local enlarged figure), but the interface reaction resistance and internal diffusion resistance increase with the rise of temperature in the later stage. It can be explained by the following analysis: From Equation \ref{e2-12} and \ref{e2-13}, it can be seen that heating increases the reaction rate constant, the reduction degree of $\rm{Fe_3O_4}$ and FeO, and the equilibrium constant. The positive correlation between $A_\text{mw}$ and temperature indicates that the reduction degree and equilibrium constant of $\rm{Fe_3O_4}$ play a decisive role in the change of $A_\text{mw}$. The temperature at the early stage of the reaction has an inhibitory effect on $A_\text{wFe}$, indicating that the reaction rate constant is the main factor affecting $A_\text{wFe}$. At the later stage of the reaction, the increase of reduction degree became the dominant factor for the increase of resistance. In Figure \ref{fig17} (c) and (d), the influence of temperature on diffusion resistance is similar to that of interface reaction resistance, which can be analyzed by Equation \ref{e2-14} and will not be repeated here. In addition, the time-averaged value of gas film diffusion resistance at 1073 K and 973 K are 0.14 and 0.12 (s/m), respectively. Therefore, the reduction of the hematite particles is also controlled by the interfacial chemical reaction, just like magnetite powder in Section 5.1.

\section{Conclusion}\label{s6}
The direct reduction of iron ore in hydrogen was simulated by using CPU-GPU hybrid parallel coarse-grained CFD-DEM-IBM method. Based on the USCM particle reaction model, a variety of resistance networks are considered in the modeling to reflect the real single particle reaction process. The kinetic parameters (pre-exponential factor) of each sub-reaction are determined by comparing the simulated and experimental values of the particle overall reduction. Heat transfer, mass transfer and chemical reaction behavior of magnetite and hematite particles in fluidized beds at high temperature were predicted. The following conclusions were drawn: (i) The CFD-DEM method coupled with USCM model can accurately predict the step-by-step reduction process of iron ore particles at high temperature; (ii) The interfacial chemical reaction of hematite and magnetite particles used in the simulation during reduction is a rate control step, and the interfacial reaction resistance $>$ diffusion resistance of gas through product layer $>$ gas film resistance, so appropriate reaction kinetic parameters is the key to accurately predict iron ore reduction; (iii) Heat is released when hematite is reduced to magnetite in hydrogen, and high temperature will inhibit the generation of magnetite, while the reduction of magnetite is an endothermic process, and rising temperature is conducive to the generation of iron; (iv) The complete conversion time of fine magnetite powder is much less than that of medium sized hematite particles.
Clearly, the present work provides a computational tool for the rapid and efficient simulation of iron ore reduction in fluidized beds by combining the particle scale reaction model with the traditional CFD-DEM method, and reveals the reaction control mechanism and the basic principles of momentum, mass and energy transfer during fluidization reduction. It can be applied to the optimization and design of industrial reactors with complex geometric structures and large particle numbers, not only in fluidized beds, but also in metallurgical equipment such as blast furnaces.

\section*{CRediT authorship contribution statement}
$\textbf{Bin Lan}$: Methodology, Software, Formal analysis, Investigation, Writing - original draft. $\textbf{Ji Xu}$: Software, Writing - review $\&$  editing. $\textbf{Shuai Lu}$: Formal analysis, Writing - review $\&$  editing. $\textbf{Yige Liu}$: Software, Writing - review $\&$  editing. $\textbf{Fan Xu}$: Formal analysis, Writing - review $\&$  editing. $\textbf{Bidan Zhao}$: Funding acquisition, Writing - review $\&$  editing. $\textbf{Zheng Zou}$: Funding acquisition, Writing - review $\&$  editing. $\textbf{Ming Zhai}$: Supervision, Writing - review $\&$ editing. $\textbf{Junwu Wang}$: Methodology, Conceptualization, Funding acquisition, Supervision, Writing - review $\&$  editing.

\section*{Declaration of Competing Interest}
The authors declare that they have no known competing financial interests or personal relationships that could have appeared to influence the work reported in this paper.

\section*{Acknowledgement}
This study is financially supported by the Strategic Priority Research Program of the Chinese Academy of Sciences (XDA29040200), the National Natural Science Foundation of China (21978295, 22311530057, 11988102), the Innovation Academy for Green Manufacture, Chinese Academy of Sciences (IAGM2022D02), the National Key R$\&$D Program of China (2021YFB1715500), the Young Elite Scientists Sponsorship Program by CAST (2022QNRC001) and the Youth Innovation Promotion Association, Chinese Academy of Sciences under Grant No.2019050.

\appendix
\section{Correlation coefficients of species thermophysical properties}
\renewcommand\thetable{\Alph{section}\arabic{table}}
\setcounter{table}{0}
\begin{table}[]
\centering
\caption{Correlation coefficients of the standard enthalpy with temperature.}\label{tab c1}
\begin{tabular}{cccccccc}
\hline
Species & Temperature (K)  & $a_\text{1}$  &$a_\text{2}$ &$a_\text{3}$ &$a_\text{4}$ &$a_\text{5}$ &$a_\text{6}$  \\ \hline
\multirow{2}{*}{$\rm{H_2}$}      & 298$\sim$1000    & 2.3443   & 7.98e-3 & -1.95e-5 & 2.02e-8 &-7.38e-12   & -9.179e2    \\
                & 1000$\sim$6000      &2.9328  & 8.27e-4     &  -1.46e-7          &1.54e-11        & -6.89e-16     &  -8.131e2          \\
\multirow{2}{*}{$\rm{H_2O}$}    &373$\sim$1000   &4.1986  & -2.04e-3  & 6.52e-6          &-5.49e-9    &1.77e-12   &  -3.03e4      \\
                & 1000$\sim$6000     &2.677  &2.97e-3  & -7.74e-7           & 9.44e-11    & -4.27e-15   & -2.989e4       \\
\multirow{2}{*}{$\rm{N_2}$}   & 298$\sim$1000   & 3.531  & -1.24e-4    & -5.03e-7           &  2.44e-9    & -1.41e-12    &  -1.047e3        \\
                 &1000$\sim$6000   &2.9526  & 1.40e-3  & -4.93e-7          & 7.86e-11   & -4.61e-15  &   -9.239e2     \\
\multirow{2}{*}{$\rm{Fe_2O_3}$}   & 298$\sim$960  &0.1522  & 6.71e-2  &  -1.13e-4         & 9.93e-8   & -3.28e-11  & -1.013e5       \\
                & 960$\sim$1700     & 353.05 &-0.9728  & 1.05e-3           & -4.96e-7    &8.74e-11    & -1.96e5       \\
\multirow{2}{*}{$\rm{Fe_3O_4}$}    & 298$\sim$850   & 4.8445  &  4.39e-2   &  5.25e-5          &-2.21e-7      &1.75e-10     &-1.38e5          \\
                & 850$\sim$1870     & 88.431 & -0.149 &  1.26e-4          & -4.7e-8    & 6.79e-12   & -1.621e5       \\
\multirow{2}{*}{$\rm{FeO}$}   &298$\sim$1000   & 5.3195 &2.21e-3   & 1.07e-6          &-2.79e-9	&1.33e-12	&-3.441e4   \\
                 &1000$\sim$1650   & 5.8316	& 1.43e-3	& -9.32e-8	& -6.6e-12	& -2.25e-14	& -3.457e4       \\
\multirow{4}{*}{$\rm{Fe}$}   &298$\sim$1000         & 2.4134	 & -1.58e-3	 & 2.15e-5	 & -3.8e-8	 & 2.20e-11	 & -774.38  \\
                 &1000$\sim$1042     & 4690.8	 & -9.91	 & 2.69e-3	 & 5.54e-6	 & -3.02e-9	 & -1.415e6      \\
                &1042$\sim$1184      & 659.68	 & -1.14	 & 4.96e-4	 & 0	 & 0	 & -2.52e5     \\
                &1184$\sim$1665       & 61.011	 & -0.161	 & 1.68e-4	 & -7.75e-8	 & 1.33e-11	 & -1.65e4   \\   \hline
\end{tabular}
\end{table}

\begin{table}[]
\centering
\caption{Correlation coefficients of the gas viscosities with temperature.}\label{tab c2}
\begin{tabular}{ccccc}
\hline
Species & Temperature (K)  & $a_\text{1}$  &$a_\text{2}$ &$a_\text{3}$   \\ \hline
$\rm{H_2}$ & 298$\sim$2000      &2.85e-6  & 2.16e-8 & -3.78e-12             \\
$\rm{H_2O}$ & 373$\sim$1350 &-4.44e-6	&4.52e-8	&-3.20e-12   \\
$\rm{N_2}$ & 298$\sim$2000 &7.86e-6	&3.87e-8	&-5.11e-12   \\ \hline
\end{tabular}
\end{table}

\begin{table}[]
\centering
\caption{Correlation coefficients of the gas thermal conductivity coefficients with temperature.}\label{tab c3}
\begin{tabular}{ccccccc}
\hline
Species & Temperature (K)  & $a_\text{1}$  &$a_\text{2}$ &$a_\text{3}$ &$a_\text{4}$ &$a_\text{5}$  \\ \hline
$\rm{H_2}$ & 298$\sim$2000    &3.07e-3	&8.09e-4	&-8.52e-7	&7.2e-10	&-2.21e-13             \\
$\rm{H_2O}$ & 373$\sim$1350   &6.84e-3	&6.25e-6	&1.31e-7	&-5.99e-11	&1.12e-14 \\
$\rm{N_2}$ & 298$\sim$2000    &2.51e-3	&8.92e-5	&-4.00e-8	&1.62e-11	&-2.61e-15\\ \hline
\end{tabular}
\end{table}



\begin{thebibliography}{10}

\bibitem{zhou2016emission}
K.~Zhou, S.~Yang, Emission reduction of {C}hina's steel industry: Progress and
  challenges, Renewable and Sustainable Energy Reviews 61 (2016) 319--327.

\bibitem{dutta2020basic}
S.~K. Dutta, Y.~B. Chokshi, Basic concepts of Iron and steel making, Springer
  Nature, 2020.

\bibitem{soni2022review}
R.~K. Soni, E.~Chinthapudi, S.~K. Tripathy, M.~Bose, P.~S. Goswami, Review on
  the chemical reduction modelling of hematite iron ore to magnetite in
  fluidized bed reactor, Reviews in Chemical Engineering (2022) 1--44.

\bibitem{lu2015quality}
L.~Lu, J.~Pan, D.~Zhu, Quality requirements of iron ore for iron production,
  in: Iron Ore, Elsevier, 2015, pp. 475--504.

\bibitem{sun2014gas}
T.~Sun, Y.~Shen, J.~Jia, Gas cleaning and hydrogen sulfide removal for corex
  coal gas by sorption enhanced catalytic oxidation over recyclable activated
  carbon desulfurizer, Environmental science \& technology 48~(4) (2014)
  2263--2272.

\bibitem{yi2019finex}
S.-H. Yi, M.-E. Choi, D.-H. Kim, C.-K. Ko, W.-I. Park, S.-Y. Kim,
  Finex{\textregistered} as an environmentally sustainable ironmaking process,
  Ironmaking \& Steelmaking 46~(7) (2019) 625--631.

\bibitem{jeong2015system}
S.-J. Jeong, System dynamics approach for the impacts of finex technology and
  carbon taxes on steel demand: Case study of the posco, International Journal
  of Precision Engineering and Manufacturing-Green Technology 2 (2015) 85--93.

\bibitem{kwauk2007handbook}
M.~Kwauk, H.~Li, Handbook of fluidization, Chemical Industry Press Beijing,
  2008.

\bibitem{he2017direct}
S.~He, H.~Sun, C.~Hu, J.~Li, Q.~Zhu, H.~Li, Direct reduction of fine iron ore
  concentrate in a conical fluidized bed, Powder Technology 313 (2017)
  161--168.

\bibitem{brown1966fior}
J.~Brown, D.~Campbell, A.~Saxton, J.~Carr~Jr, {FIOR}-the {E}sso fluid iron ore
  direct reduction process, JOM 18~(2) (1966) 237--242.

\bibitem{battle2014direct}
T.~Battle, U.~Srivastava, J.~Kopfle, R.~Hunter, J.~McClelland, The direct
  reduction of iron, in: Treatise on process metallurgy, Elsevier, 2014, pp.
  89--176.

\bibitem{plaul2009fluidized}
F.~Plaul, C.~B{\"o}hm, J.~Schenk, Fluidized-bed technology for the production
  of iron products for steelmaking, Journal of the Southern African Institute
  of Mining and Metallurgy 109~(2) (2009) 121--128.

\bibitem{yu2010application}
Y.~Yu, W.~Chen, Application of flash magnetizing roasting technique in
  beneficiation of siderite and limonite, in: The 2010 international symposium
  on project management, 2010, pp. 13--17.

\bibitem{yu2019growth}
J.-w. Yu, Y.-x. Han, Y.-j. Li, P.~Gao, Growth behavior of the magnetite phase
  in the reduction of hematite via a fluidized bed, International Journal of
  Minerals, Metallurgy, and Materials 26 (2019) 1231--1238.

\bibitem{li2015phase}
Y.-j. Li, R.~Wang, Y.-x. Han, X.-c. Wei, Phase transformation in suspension
  roasting of oolitic hematite ore, Journal of Central South University 22~(12)
  (2015) 4560--4565.

\bibitem{du2016role}
Z.~Du, Q.~Zhu, Y.~Yang, C.~Fan, F.~Pan, H.~Sun, Z.~Xie, The role of {M}g{O}
  powder in preventing defluidization during fluidized bed reduction of fine
  iron ores with different iron valences, steel research international 87~(12)
  (2016) 1742--1749.

\bibitem{du2022effect}
Z.~Du, Y.~Ge, F.~Liu, C.~Fan, F.~Pan, Effect of different modification methods
  on fluidized bed hydrogen reduction of cohesive iron ore fines, Powder
  Technology 400 (2022) 117226.

\bibitem{du2022relationship}
Z.~Du, J.~Liu, F.~Liu, F.~Pan, Relationship of particle size, reaction and
  sticking behavior of iron ore fines toward efficient fluidized bed reduction,
  Chemical Engineering Journal 447 (2022) 137588.

\bibitem{du2017enhanced}
Z.~Du, Q.~Zhu, C.~Fan, F.~Pan, Z.~Xie, Enhanced effect and mechanism of
  {F}e2{O}3 on cao for defluidization inhibition during fluidized bed reduction
  of iron ore fines, Powder Technology 313 (2017) 82--87.

\bibitem{spreitzer2019iron}
D.~Spreitzer, J.~Schenk, Iron ore reduction by hydrogen using a laboratory
  scale fluidized bed reactor: Kinetic investigation—experimental setup and
  method for determination, Metallurgical and materials transactions B 50
  (2019) 2471--2484.

\bibitem{spreitzer2020fluidization}
D.~Spreitzer, J.~Schenk, Fluidization behavior and reducibility of iron ore
  fines during hydrogen-induced fluidized bed reduction, Particuology 52 (2020)
  36--46.

\bibitem{zhong2018model}
Y.~Zhong, J.~Gao, Z.~Guo, A model for solid surface viscosity of iron powders
  at high temperature: Influence of particle size distribution, Powder
  Technology 335 (2018) 371--374.

\bibitem{komatina2018sticking}
M.~Komatina, H.~W. GUDENAU, The sticking problem during direct reduction of
  fine iron ore in the fluidized bed, Metallurgical and Materials Engineering.

\bibitem{lei2016optimization}
C.~Lei, G.~Zhang, Q.~Zhu, Z.~Xie, Optimization of carbon deposition process
  during the pre-reduction of fine iron ore in a fluidized bed, Powder
  Technology 296 (2016) 79--86.

\bibitem{lei2014experimental}
C.~Lei, Q.~Zhu, H.~Li, Experimental and theoretical study on the fluidization
  behaviors of iron powder at high temperature, Chemical Engineering Science
  118 (2014) 50--59.

\bibitem{zhong2012defluidization}
Y.~Zhong, Z.~Wang, Z.~Guo, Q.~Tang, Defluidization behavior of iron powders at
  elevated temperature: Influence of fluidizing gas and particle adhesion,
  Powder Technology 230 (2012) 225--231.

\bibitem{holappa2017recent}
L.~Holappa, Recent achievements in iron and steel technology., Journal of
  Chemical Technology \& Metallurgy 52~(2) (2017) 159--167.

\bibitem{an2018potential}
R.~An, B.~Yu, R.~Li, Y.-M. Wei, Potential of energy savings and {CO}2 emission
  reduction in {C}hina's iron and steel industry, Applied energy 226 (2018)
  862--880.

\bibitem{patisson2020hydrogen}
F.~Patisson, O.~Mirgaux, Hydrogen ironmaking: How it works, Metals 10~(7)
  (2020) 922.

\bibitem{shi2005modelling}
J.~Shi, E.~Donskoi, D.~S. McElwain, L.~J. Wibberley, Modelling the reduction of
  an iron ore-coal composite pellet with conduction and convection in an
  axisymmetric temperature field, Mathematical and computer modelling 42~(1-2)
  (2005) 45--60.

\bibitem{valipour2009mathematical}
M.~Valipour, Mathematical modeling of a non-catalytic gas-solid reaction:
  hematite pellet reduction with syngas, Scientia Iranica 16~(2) (2009)
  108--124.

\bibitem{tang2012simulation}
H.~Tang, Z.~Guo, K.~Kitagawa, Simulation study on performance of z-path
  moving-fluidized bed for gaseous reduction of iron ore fines, ISIJ
  international 52~(7) (2012) 1241--1249.

\bibitem{natsui2014numerical}
S.~Natsui, T.~Kikuchi, R.~O. Suzuki, Numerical analysis of carbon
  monoxide--hydrogen gas reduction of iron ore in a packed bed by an
  euler--lagrange approach, Metallurgical and Materials Transactions B 45
  (2014) 2395--2413.

\bibitem{nouri2011simulation}
S.~Nouri, H.~A. Ebrahim, E.~Jamshidi, Simulation of direct reduction reactor by
  the grain model, Chemical Engineering Journal 166~(2) (2011) 704--709.

\bibitem{ariyan2016numerical}
Z.~G. Ariyan, V.~Mohammad~Sadegh, B.~Mojtaba, Numerical analysis of complicated
  heat and mass transfer inside a wustite pellet during reducing to sponge iron
  by {H}2 and {CO} gaseous mixture, Journal of Iron and Steel Research
  International 23~(11) (2016) 1142--1150.

\bibitem{kinaci2018direct}
M.~E. Kinaci, T.~Lichtenegger, S.~Schneiderbauer, Direct reduction of iron-ore
  in fluidized beds, in: Computer Aided Chemical Engineering, Vol.~43,
  Elsevier, 2018, pp. 217--222.

\bibitem{kinaci2020cfd}
M.~Kinaci, T.~Lichtenegger, S.~Schneiderbauer, A {CFD-DEM} model for the
  simulation of direct reduction of iron-ore in fluidized beds, Chemical
  Engineering Science 227 (2020) 115858.

\bibitem{schneiderbauer2020computational}
S.~Schneiderbauer, M.~E. Kinaci, F.~Hauzenberger, Computational fluid dynamics
  simulation of iron ore reduction in industrial-scale fluidized beds, steel
  Research International 91~(12) (2020) 2000232.

\bibitem{wan2022numerical}
Z.-w. Wan, J.-y. Huang, G.-m. Zhu, Q.-y. Xu, Numerical simulation of the
  operating conditions for the reduction of iron ore powder in a fluidized bed
  based on the {CPFD} method, Processes 10~(9) (2022) 1870.

\bibitem{rosser2023investigation}
J.~G. Rosser, Investigation of jamming phenomenon in a {DRI} furnace pellet
  feed system using the discrete element method and computational fluid
  dynamics, Ph.D. thesis, Purdue University Graduate School (2023).

\bibitem{dianyu2020validation}
E.~Dianyu, Validation of {CFD-DEM} model for iron ore reduction at particle
  level and parametric study, Particuology 51 (2020) 163--172.

\bibitem{xu2022coarse}
D.~Xu, S.~Wang, Y.~Shen, Coarse-grained r{CFD-DEM} analysis of coke
  gasification and iron ore reduction in the shaft region of an ironmaking
  blast furnace, Powder Technology 408 (2022) 117706.

\bibitem{levenspiel1998chemical}
O.~Levenspiel, Chemical {R}eaction {E}ngineering, John {W}iley \& {S}ons, 1998.

\bibitem{kinaci2017modelling}
M.~E. Kinaci, T.~Lichtenegger, S.~Schneiderbauer, Modelling of chemical
  reactions in metallurgical processes, in: Proceedings of the 12th
  International Conference on Computational Fluid Dynamics in the Oil \& Gas,
  Metallurgical and Process Industries, SINTEF Academic Press, 2017.

\bibitem{spreitzer2019reduction}
D.~Spreitzer, J.~Schenk, Reduction of iron oxides with hydrogen-a review, Steel
  Research International 90~(10) (2019) 1900108.

\bibitem{spitzer1966generalized}
R.~Spitzer, F.~Manning, W.~Philbrook, Generalized model for the gaseous,
  topochemical reduction of porous hematite spheres, AIME Met Soc Trans
  236~(12) (1966) 1715--1724.

\bibitem{zhang2013thermodynamic}
W.~Zhang, J.~Zhang, Q.~Li, Y.~He, B.~Tang, M.~Li, Z.~Zhang, Z.~Zou,
  Thermodynamic analyses of iron oxides redox reactions, in: Proceedings of the
  8 th Pacific Rim International Congress on Advanced Materials and Processing,
  Springer, 2013, pp. 777--789.

\bibitem{takenaka1986mathematical}
Y.~Takenaka, Y.~Kimura, K.~Narita, D.~Kaneko, Mathematical model of direct
  reduction shaft furnace and its application to actual operations of a model
  plant, Computers \& chemical engineering 10~(1) (1986) 67--75.

\bibitem{Gunn1978}
D.~Gunn, Transfer of heat or mass to particles in fixed and fluidised beds,
  International Journal of Heat and Mass Transfer 21~(4) (1978) 467--476.

\bibitem{li2017reaction}
S.~Li, F.~Xin, L.~Li, Reaction Engineering, Butterworth-Heinemann, 2017.

\bibitem{fuller1966new}
E.~N. Fuller, P.~D. Schettler, J.~C. Giddings, New method for prediction of
  binary gas-phase diffusion coefficients, Industrial \& Engineering Chemistry
  58~(5) (1966) 18--27.

\bibitem{wilke1955correlation}
C.~Wilke, P.~Chang, Correlation of diffusion coefficients in dilute solutions,
  AIChE Journal 1~(2) (1955) 264--270.

\bibitem{wakao1962diffusion}
N.~Wakao, J.~Smith, Diffusion in catalyst pellets, Chemical Engineering Science
  17~(11) (1962) 825--834.

\bibitem{lan2020long}
B.~Lan, J.~Xu, P.~Zhao, Z.~Zou, Q.~Zhu, J.~Wang, Long-time coarse-grained
  {CFD-DEM} simulation of residence time distribution of polydisperse particles
  in a continuously operated multiple-chamber fluidized bed, Chemical
  Engineering Science 219 (2020) 115599.

\bibitem{gidaspow1994multiphase}
D.~Gidaspow, Multiphase flow and fluidization: continuum and kinetic theory
  descriptions, Academic press, 1994.

\bibitem{Ganser1993A}
G.~H. Ganser, A rational approach to drag prediction of spherical and
  nonspherical particles, Powder Technology 77~(2) (1993) 143--152.

\bibitem{Hua2015Eulerian}
L.~Hua, H.~Zhao, J.~Li, J.~Wang, Q.~Zhu, {Eulerian-Eulerian} simulation of
  irregular particles in dense gas-solid fluidized beds, Powder Technology 284
  (2015) 299--311.

\bibitem{lan2022cfd}
B.~Lan, P.~Zhao, J.~Xu, B.~Zhao, M.~Zhai, J.~Wang, {CFD-DEM-IBM} simulation of
  particle drying processes in gas-fluidized beds, Chemical Engineering Science
  255 (2022) 117653.

\bibitem{xu2019virtual}
J.~Xu, X.~Liu, S.~Hu, W.~Ge, Virtual process engineering on a three-dimensional
  circulating fluidized bed with multiscale parallel computation, Journal of
  Advanced Manufacturing and Processing 1~(1-2) (2019) e10014.

\bibitem{batchelor1977thermal}
G.~K. Batchelor, R.~O'brien, Thermal or electrical conduction through a
  granular material, Proceedings of the Royal Society of London. A.
  Mathematical and Physical Sciences 355~(1682) (1977) 313--333.

\bibitem{rong1999simulation}
D.~Rong, M.~Horio, {DEM} simulation of char combustion in a fluidized bed, in:
  Second International Conference on CFD in the Minerals and Process Industries
  CSIRO, Melbourne, Australia, 1999, pp. 65--70.

\bibitem{zhao2020Acomputational}
P.~Zhao, J.~Xu, X.~Liu, W.~Ge, J.~Wang, A computational fluid dynamics-discrete
  element-immersed boundary method for cartesian grid simulation of heat
  transfer in compressible gas--solid flow with complex geometries, Physics of
  Fluids 32~(10) (2020) 103306.

\bibitem{zhou2009particle}
Z.~Zhou, A.~Yu, P.~Zulli, Particle scale study of heat transfer in packed and
  bubbling fluidized beds, AIChE Journal 55~(4) (2009) 868--884.

\bibitem{musser2015constitutive}
J.~Musser, M.~Syamlal, M.~Shahnam, D.~Huckaby, Constitutive equation for heat
  transfer caused by mass transfer, Chemical Engineering Science 123 (2015)
  436--443.

\bibitem{burcat2005third}
A.~Burcat, B.~Ruscic, Third millenium ideal gas and condensed phase
  thermochemical database for combustion (with update from active
  thermochemical tables)., Tech. rep., Argonne National Lab.(ANL), Argonne, IL
  (United States) (2005).

\bibitem{mehrabadi2015pseudo}
M.~Mehrabadi, S.~Tenneti, R.~Garg, S.~Subramaniam, Pseudo-turbulent gas-phase
  velocity fluctuations in homogeneous gas--solid flow: fixed particle
  assemblies and freely evolving suspensions, Journal of Fluid Mechanics 770
  (2015) 210--246.

\bibitem{peng2019implementation}
C.~Peng, B.~Kong, J.~Zhou, B.~Sun, A.~Passalacqua, S.~Subramaniam, R.~Fox,
  Implementation of pseudo-turbulence closures in an eulerian--eulerian
  two-fluid model for non-isothermal gas--solid flow, Chemical Engineering
  Science 207 (2019) 663--671.

\bibitem{sun2015modeling}
B.~Sun, S.~Tenneti, S.~Subramaniam, Modeling average gas--solid heat transfer
  using particle-resolved direct numerical simulation, International Journal of
  Heat and Mass Transfer 86 (2015) 898--913.

\bibitem{norouzi2016coupled}
H.~R. Norouzi, R.~Zarghami, R.~Sotudeh-Gharebagh, N.~Mostoufi, Coupled CFD-DEM
  modeling: formulation, implementation and application to multiphase flows,
  John Wiley \& Sons, 2016.

\bibitem{Lu2014EMMS}
L.~Lu, J.~Xu, W.~Ge, Y.~Yue, X.~Liu, J.~Li, {EMMS}-based discrete particle
  method ({EMMS}-{DPM}) for simulation of gas-solid flows, Chemical Engineering
  Science 120 (2014) 67--87.

\bibitem{lu2016computer}
L.~Lu, J.~Xu, W.~Ge, G.~Gao, Y.~Jiang, M.~Zhao, X.~Liu, J.~Li, Computer virtual
  experiment on fluidized beds using a coarse-grained discrete particle
  method-{EMMS-DPM}, Chemical Engineering Science 155 (2016) 314--337.

\bibitem{lan2021simulation}
B.~Lan, J.~Xu, Z.~Liu, J.~Wang, Simulation of scale-up effect of particle
  residence time distribution characteristics in continuously operated
  dense-phase fluidized beds, CIESC Journal 72 (2021) 521--533.

\bibitem{lu2017extension}
L.~Lu, A.~Morris, T.~Li, S.~Benyahia, Extension of a coarse grained particle
  method to simulate heat transfer in fluidized beds, International Journal of
  Heat and Mass Transfer 111 (2017) 723--735.

\bibitem{hu2019advances}
C.~Hu, Y.~He, D.~Liu, S.~Sun, D.~Li, Q.~Zhu, J.~Yu, Advances in mineral
  processing technologies related to iron, magnesium, and lithium, Reviews in
  Chemical Engineering 36~(1) (2019) 107--146.

\bibitem{yu2021coarse}
J.~Yu, L.~Lu, Y.~Xu, X.~Gao, M.~Shahnam, W.~Rogers, Coarse-grained {CFD-DEM}
  simulation and the design of an industrial-scale coal gasifier, Industrial \&
  Engineering Chemistry Research 61~(1) (2021) 866--881.

\bibitem{du2023coarse}
S.~Du, J.~Wang, Y.~Yu, Q.~Zhou, Coarse-grained {CFD-DEM} simulation of coal and
  biomass co-gasification process in a fluidized bed reactor: Effects of
  particle size distribution and operating pressure, Renewable Energy 202
  (2023) 483--498.

\bibitem{doble2007perry}
M.~Doble, Perry's {C}hemical {E}ngineers' {H}andbook, McGraw-Hil, New York, US.

\bibitem{niksiar2010design}
A.~Niksiar, A.~Rahimi, Design of a moving bed reactor for the production of
  uranium tetrafluoride based on mathematical modeling, Chemical engineering
  science 65~(10) (2010) 3147--3157.

\bibitem{zhao2020cfd}
P.~Zhao, J.~Xu, W.~Ge, J.~Wang, A {CFD-DEM-IBM method for Cartesian} grid
  simulation of gas-solid flow in complex geometries, Chemical Engineering
  Journal 389 (2020) 124343.

\bibitem{zhao2022cartesian}
P.~Zhao, J.~Xu, B.~Zhao, D.~Li, J.~Wang, {Cartesian grid simulation of reacting
  gas-solid flow using CFD-DEM-IBM} method, Powder Technology 407 (2022)
  117651.

\bibitem{zhao2022euler}
P.~Zhao, J.~Xu, Q.~Chang, W.~Ge, J.~Wang, {Euler-Lagrange} simulation of dense
  gas-solid flow with local grid refinement, Powder Technology 399 (2022)
  117199.

\bibitem{lan2023critical}
B.~Lan, P.~Zhao, J.~Xu, B.~Zhao, M.~Zhai, J.~Wang, The critical role of scale
  resolution in {CFD} simulation of gas-solid flows: A heat transfer study
  using {CFD-DEM-IBM} method, Chemical Engineering Science 266 (2023) 118268.

\bibitem{li2021direct}
S.~Li, P.~Zhao, J.~Xu, L.~Zhang, J.~Wang, Direct comparison of {CFD-DEM
  simulation and experimental measurement of Geldart A} particles in a
  micro-fluidized bed, Chemical Engineering Science 242 (2021) 116725.

\bibitem{li2022cfd}
S.~Li, P.~Zhao, J.~Xu, L.~Zhang, J.~Wang, {CFD-DEM simulation of polydisperse
  gas-solid flow of Geldart A} particles in bubbling micro-fluidized beds,
  Chemical Engineering Science 253 (2022) 117551.

\bibitem{heidari2021review}
A.~Heidari, N.~Niknahad, M.~Iljana, T.~Fabritius, A review on the kinetics of
  iron ore reduction by hydrogen, Materials 14~(24) (2021) 7540.

\bibitem{nicolle1979mechanism}
R.~Nicolle, A.~Rist, The mechanism of whisker growth in the reduction of
  w{\"u}stite, Metallurgical Transactions B 10 (1979) 429--438.

\bibitem{spreitzer2020development}
D.~Spreitzer, Development of characterization methods for the evaluation of the
  kinetic behavior and the fluidization of iron ore fines during
  hydrogen-induced fluidized bed reduction, Ph.D. thesis, Montan University
  (2020).

\end{thebibliography}

\end{spacing}
\end{document}